\documentclass{aa}

\newcommand{\colasolver}{\texttt{COLASolver}}

\newcommand{\sesame}{\textit{Sesame}}

\usepackage{bm}

\usepackage{xcolor}

\newcommand{\lcdm}{$\Lambda$CDM}
\newcommand{\fofr}{$f(R)$}
\usepackage{graphicx}
\usepackage{mathrsfs}
\usepackage{float}
\usepackage{txfonts}
\definecolor{darkblue}{rgb}{0.0,0.0,0.55}
\usepackage{hyperref}
\hypersetup{
    colorlinks = true,
    linkcolor = darkblue,
    anchorcolor = darkblue,
    citecolor = darkblue,
    filecolor = darkblue,
    urlcolor = darkblue
    }

\begin{document}
   \title{\sesame: A power spectrum emulator pipeline for beyond-$\Lambda$CDM models}
    
   \author{Renate Mauland\inst{1}
          \and
          Hans A. Winther\inst{1}
          \and
          Cheng-Zong Ruan\inst{1}
          }

   \institute{Institute of Theoretical Astrophysics, University of Oslo, P.O.Box 1029 Blindern, 0315 Oslo, Norway\\
              \email{renate.mauland-hus@astro.uio.no}
             }

   \date{Received ---; accepted ---}

  \abstract{The mysterious nature of the dark sector of the $\Lambda$-cold-dark-matter ($\Lambda$CDM) model is one of the main motivators behind the study of alternative cosmological models. A central quantity of interest for these models is the matter power spectrum, which quantifies structure formation on various scales and can be cross-validated through theory, simulations, and observations. Here, we present a tool that can be used to create emulators for the non-linear matter power spectrum, and similar global clustering statistics, for models beyond $\Lambda$CDM with very little computation effort and without the need for supercomputers. We use fast approximate \textit{N}-body simulations to emulate the boost, $B(k,z)=P_{\text{beyond}-\Lambda \rm CDM}(k,z) / P_{\Lambda \rm CDM}(k, z)$, and then rely on existing high-quality emulators made for $\Lambda$CDM to isolate $P_{\text{beyond}-\Lambda \rm CDM}(k,z)$. Since both the $\Lambda$CDM and beyond-$\Lambda$CDM models are simulated in our approach, some of the lack of power on small scales due to the low force-resolution in the simulations is factored out, allowing us to extend the emulator to $k \sim 3-5\,h\,\mathrm{Mpc}^{-1}$ and still maintain good accuracy. In addition, errors from the simulation and emulation process can easily be estimated and factored into the covariance when using the emulator on data. As an example of using the pipeline, we create an emulator for the well-studied $f(R)$ model with massive neutrinos, using approximately $3000$ CPU hours of computation time. Provided with the paper is a fully functioning pipeline that generates parameter samples, runs a Boltzmann solver to produce initial conditions, runs the simulations, and then gathers all the data and runs it through a machine learning module to develop the emulator. This tool, named \sesame, can be used by anyone to generate a power spectrum emulator for the cosmological model of their choice.   
  }


   \maketitle
%
\section{Introduction}
The $\Lambda$-cold-dark-matter (\lcdm) model describes our Universe well, yet two of its main components remain elusive. The true natures of dark matter and dark energy are still unknown, but their impact on the Universe has been, and continues to be, widely studied across multiple research fields. In an attempt to forego the dark energy component of the \lcdm{} model, alternative theories of gravity have become a popular avenue to explore. These beyond-\lcdm{} models \citep[see e.g.][]{Clifton2012,Koyama2016,Wright2023} have an effect on structure formation, leaving an imprint on the matter power spectrum. This can be further studied through the use of cosmological simulations, which typically require a large amount of computing resources for high-resolution simulations capable of accurately distinguishing between models down to small scales. In addition, a simulation is only performed for a specified set of cosmological parameters, requiring a rerun for any parameter changes. To forgo both of these issues, emulators can be created for desired statistical observables, like the matter power spectrum - a key observable whose theoretical prediction is needed to constrain beyond-\lcdm{} models in current and near-future weak-lensing surveys \citep{Benitez2014,Ivezic2019,DES2021,Scaramella2022}. The emulators \citep{Heitmann2013,Kwan2015,Giblin2019,Nishimichi2019,Angulo2021,knabenhans2021,Moran2023} are typically constructed by performing a high number of \textit{N}-body simulations within some parameter space, and then interpolating to access any desired parameter value. This can be done, for example, through the use of machine learning, training a neural network on highly accurate simulation data. As mentioned above, the simulation step can be computationally expensive, but once it is performed and the following training is done, the emulators are simple to use and have both minimal time and memory requirements. 

Although highly accurate, a limit of this approach is the ability to easily extend the resulting emulator to new cosmological models. In this paper, we present a full pipeline using fast approximate \textit{N}-body simulations and neural network training to create an emulator for the matter power spectrum boost, $B(k,z) = P_{\text{beyond}-\Lambda \rm CDM}(k,z) / P_{\Lambda \rm CDM}(k, z)$. The approximate simulations employ the comoving Lagrangian acceleration (COLA) method \citep{Tassev2013} to simulate both the \lcdm{} and beyond-\lcdm{} models \citep{Valogiannis2017,Winther2017,Wright2017,Brando2022,Fiorini2022,Brando2023,Wright2023}, allowing us to extract the boost up to scales of $k \sim 3-5\,h\,\mathrm{Mpc}^{-1}$ to a few percent accuracy. The pipeline is named \sesame{} - from \textit{s}imulations to \textit{e}mulator\textit{s} using \textit{a}pproximate \textit{me}thods. As a demonstration of \sesame, we create an emulator for the boost between the Hu-Sawicki \fofr{} model \citep{Hu2007} and a dynamical dark energy model, $w_0w_a$CDM. In \fofr-modified gravity, an additional function of the Ricci scalar, $R$, is added to the general relativity (GR) framework \citep{Buchdahl1970}. This function can be designed to recreate a similar expansion history as \lcdm, without the need for dark energy. Still, as the nature of gravity is modified, resulting observational signals are expected \citep[see e.g.][for a detailed review]{deFelice2010}. One such signal is the enhancement of structure formation on scales smaller than the Compton wavelength of the scalaron - the scalar degree of freedom of the \fofr{} theory, $\mathrm{d}f/\mathrm{d}R$ \citep[e.g.][]{Hu2007,Pogosian2008,Cataneo2015}. This shows up in the matter power spectrum. 

In addition to exploring universe models besides \lcdm, calculations and simulations within the \lcdm{} framework are continuously expanded to reach higher levels of accuracy. One such extension is the inclusion of massive neutrinos. These lightweight particles have often been excluded from cosmological simulations due to their low impact compared to cold dark matter (cdm), which makes up about 84\% \citep{Planck2018} of the matter content of the Universe. However, improvements in telescopes and satellites now give us an observational accuracy high enough to measure the impact of neutrinos on structure formation - suppression on scales smaller than the neutrino free-streaming length \citep{Lesgourgues2006}. Surveys like the newly launched \textit{Euclid} satellite \citep{Laureijs2011} and the ongoing DESI (Dark Energy Spectroscopic Instrument) experiment\footnote{https://www.desi.lbl.gov} \citep{DESI2016} can measure the effect of massive neutrinos and thereby put tighter constraints on the neutrino mass scale. Because of this, we include modified gravity, massive neutrinos, and dark energy in the form of the well-known $w_0w_a$ Chevallier-Polarski-Linder (CPL) parametrisation \citep{Chevallier2001,Linder2003} when creating our emulator. The inclusion of massive neutrinos in the \fofr{} simulations is also particularly important, due to the degeneracy between the effects of neutrinos and \fofr-modified gravity on structure formation on non-linear scales \citep[e.g.][]{Baldi2014}. 

Simulations including massive neutrinos \citep{Potter2016,Adamek2017,Liu2018,Dakin2019,Partmann2020,Weinberger2020,Springel2021,Adamek2023}, modified gravity \citep{Liecosmog2012,Puchwein2013,Llinares2014,Winther2015,Hassani2020,Ruan2022,Wright2023}, and both \citep{Baldi2014,Wright2017,Giocoli2018,Mauland2023} already exist with various methods of implementation, along with models, fits, and emulators to extract the boost or the matter power spectrum directly for these cosmological models \citep[e.g.][]{Zhao2014,Winther2019,HannestadWong2020,Bose2020,Bose2021,Ramachandra2021,knabenhans2021,Arnold2022,Bose2023,Gupta2023,Moran2023}. The main takeaway from this paper is therefore not the \fofr-modified gravity emulator (although it will be provided), but the full pipeline, \sesame, which includes the drawing of parameter samples, running the simulations, training the neural network, and creating the emulator for the boost, $B(k,z)$. This tool can be used to produce an emulator for a desired cosmological model by implementing said model into the simulations and using a suitable Boltzmann solver to extract the initial conditions. The resulting accuracy of both the simulations and the emulator can be tuned by the choice of simulation settings and neural network architecture. 

This paper is structured as follows: In Sect.~\ref{sec:theory}, we present some background theory for the matter power spectrum, \fofr-modified gravity, and massive neutrinos. This is followed by an outline of the methods applied in Sect.~\ref{sec:method}, including a description of the full pipeline. In Sect.~\ref{sec:simulations} we go through some simulation details, and then present our results in Sect.~\ref{sec:results}. Finally, we conclude in Sect.~\ref{sec:conclusions}.

\section{Theory}
\label{sec:theory}
In this section, we present some background information for the key components of this work. We first outline the necessary details on the matter power spectrum, followed by \fofr-modified gravity and massive neutrinos.

\subsection{Matter power spectrum}
\label{sec:mps}
The matter power spectrum, $P(k)$, is defined as \citep[e.g.][]{Peebles1980,Dodelson2020}: 
\begin{align}
    (2\pi)^3P(k)\delta_\mathrm{D}(\bm{k}-\bm{k}') = \langle \tilde{\delta}(\bm{k})\tilde{\delta}(\bm{k}')^*\rangle,
\end{align}
where $k$ is the wavenumber, $\delta_\mathrm{D}$ is the Dirac-delta function, and $\tilde{\delta}(\bm{k})$ is the Fourier transform of the overdensity field, $\delta(\bm{x})$. The power spectrum is the Fourier transform of the two-point correlation function, $\xi(\bm{r})$, which describes the excess probability, over random, of finding two objects separated by a distance $\bm{r}$. Analysing the matter power spectrum gives great insight into the clustering of matter at different times and scales, in addition to how variations in cosmological parameters affect structure formation. 

When studying alternative models to the concordance \lcdm{} model of our Universe, the ratio between the power spectrum in the alternative model and that of \lcdm{} holds valuable information about the deviations between them. Different components of a cosmological model, like massive neutrinos or modified gravity, have theoretically predicted impacts on the power spectrum \citep[e.g.][]{Lesgourgues2006,Song2007,Koyama2009}. As the matter power spectrum can be observed from various surveys \citep{Chabanier2019,Ivezic2019,Scaramella2022}, its shape is well known, and it can therefore be used to constrain these cosmological models. As an example in this paper, we are interested in the differences in the power spectrum between a $w_0w_a$CDM universe with GR as the gravity model and one with \fofr-modified gravity as the gravity model, both with the inclusion of massive neutrinos,
\begin{align}
    B(k,z) = \frac{P_{f(R)}(k,z\,\mid\,\Omega_\Lambda,\Omega_{\mathrm{cdm}},\Omega_{\mathrm{b}},n_s,\sigma_8^{f(R)},w_0,w_a,h,M_\nu,f_{R0})}{P_{\mathrm{GR}}(k,z\,\mid\,\Omega_\Lambda,\Omega_{\mathrm{cdm}},\Omega_{\mathrm{b}},n_s,\sigma_8,w_0,w_a,h,M_\nu)}.
\end{align}
Here, $\Omega_\Lambda$, $\Omega_{\mathrm{CDM}}$, and $\Omega_{\mathrm{b}}$ are the energy densities of dark energy, dark matter, and baryons respectively; $n_s$ is the scalar spectral index; $h$ is the \textit{Hubble} constant today; $\sigma_8$ and $\sigma_8^{f(R)}$ denote the normalisation of the linear matter power-spectra at $z=0$; $f_{R0}$ is the Hu-Sawicki \fofr-modified gravity parameter (see Sect.~\ref{sec:fofr}); $M_\nu$ denotes the sum of the neutrino masses, and $w_0$ and $w_a$ are dynamical dark energy parameters for the CPL parametrisation of the dark energy equation of state \citep{Chevallier2001,Linder2003},
\begin{align}
w = w_0 + w_a\frac{z}{1+z},    
\end{align}
 where $w_0 = -1$ and $w_a = 0$ correspond to a cosmological constant.

\subsection{Beyond-$\Lambda$CDM models}

A vast number of beyond-$\Lambda$CDM models are proposed in the literature, and not all of them can be covered here. For a review, we therefore refer the reader to \cite{Bull2016}.

The simplest models are dark energy models that mainly modify the background evolution through the \textit{Hubble} function, $H(a)$. These are the so-called quintessence models \citep{Wetterich1988} and parametrised models for the dark energy equation of state, $w(a)$, like CPL. Next in the level of complexity, we have models where the quintessence field is coupled to matter (often only dark matter), dubbed coupled-quintessence models \citep{Amendola2000}. Then we have modified gravity models, where an extra degree of freedom is introduced, giving rise to a fifth force for the full matter sector. To be able to evade local gravity constraints, these models often need a screening mechanism to hide the modifications in high-density environments where such gravity tests have been performed \citep[see e.g.][]{KhouryWeltman2004,Clifton2012,Koyama2016}. In addition to the models mentioned so far, we also have models of dark matter beyond cold dark matter \citep[e.g. axions][]{Marsh2016}, non-standard inflationary models \citep{Martin2014}, and many more. The model we use here for demonstrating how an emulator can be created using \sesame{} is a $f(R)$ modified gravity model. This is chosen as it is well known and because it is already implemented in the applied code base.

\subsubsection{\fofr-modified gravity}
\label{sec:fofr}
In \fofr-modified gravity theory \citep{Sotiriou2010}, the Einstein-Hilbert action of GR is extended by a function, \fofr{},
\begin{align}
    S = \bigg(\int \frac{R+f(R)}{16\pi G} + \mathcal{L}_m\bigg)\sqrt{-g}\mathrm{d}^4x.
\end{align}
Here, $R$ is the Ricci scalar, $G$ is the Newtonian gravitational constant, $\mathcal{L}_m$ is the matter Lagrangian density, and $g$ is the determinant of the metric tensor, $g_{\mu\nu}$. The \fofr{} function can take many forms, one of which is given by
\begin{align}
    f(R) = -m^2\frac{c_1(R/m^2)^n}{c_2(R/m^2)^n + 1},\label{eq:fofr}
\end{align}
proposed by \cite{Hu2007}. Here $c_1$, $c_2$, and $n$ are dimensionless, constant, and non-negative parameters of the model and $m^2 = H_0^2\Omega_\mathrm{cdm}$, with $H_0$ the value of the \textit{Hubble} parameter today. This \fofr{} function was designed so that cosmological tests at high redshifts yield the same results as for GR. In addition, in the limit where $c_2(R/m^2)^n \gg 1$, Eq.~(\ref{eq:fofr}) can be written as $f(R) = -m^2c_1/c_2 + \mathcal{O}((m^2/R)^n)$, showing that a cosmological constant, and thereby a similar background evolution to that of \lcdm{}, can be obtained by equating $-m^2c_1/c_2$ with $-2\Lambda$. This corresponds to a relation given by $c_1/c_2 = 6\Omega_\Lambda/\Omega_\mathrm{cdm}$ between the two parameters $c_1$ and $c_2$. The equation of motion of the scalar degree of freedom, $f_R$, of the \fofr-model is then given by
\begin{align}
    f_R \equiv \frac{\mathrm{d}f(R)}{\mathrm{d}R} \approx -n\frac{c_1}{c_2}\bigg(\frac{m^2}{R}\bigg)^{n+1}. 
\end{align}
By fixing the value of $f_{R0}$, the present-day background value of the scalar degree of freedom, an independent connection can be found for $c_1$ and $c_2$. This enables the model to be fully specified by the parameters $f_{R0}$ and $n$. We apply $n=1$ in this paper. 

From theory and simulations, the impact of this form of \fofr-modified gravity on structure formation, and thereby the matter power spectrum, can be predicted for various values of $f_{R0}$. In general, this modification to gravity enhances structure formation on small scales \citep{Hu2007,Pogosian2008,Cataneo2015}, as a result of an attractive force, dubbed the fifth force, which appears in addition to Newtonian gravity. The effects of this, in order for the theory to be compatible with observations \citep{Will2014}, are suppressed in high-density regions due to a chameleon screening effect \citep{Khoury2004,Brax2008}. The value of $f_{R0}$ controls the threshold at which the screening kicks in and recovers GR. Values above $f_{R0}\sim-10^{-5}$ are in general ruled out from cosmological observations \citep{Cataneo2015,Koyama2016}, although massive neutrinos, which have the opposite effect on structure formation, have not always been taken into account in these analyses \citep{Baldi2014}.

\subsection{Massive neutrinos}
\label{sec:mnu}
From particle physics, we know that there are three neutrino mass states, $\nu_{i}$ with $i=1,2,3$ \citep[e.g.][]{Thomson2013}. The absolute mass scale, $m_{\nu_i}$ (often shortened to $m_i$), of each state is unknown, but neutrino oscillation experiments give us constraints on the mass difference between the states \citep{PDG2022}
\begin{align}
    \Delta m_{21}^2 &= (7.53 \pm 0.18)\times10^{-5}\,\mathrm{eV}^2,\notag\\
    \Delta m_{32}^2 &= (-2.519 \pm 0.033)\times10^{-3}\,\mathrm{eV}^2\; (\mathrm{IH}),\\
    \Delta m_{32}^2 &= (2.437 \pm 0.033)\times10^{-3}\,\mathrm{eV}^2\; (\mathrm{NH}),\notag
    \end{align}
where IH denotes the inverted hierarchy ($m_3 \ll m_1 < m_2$) and NH the normal hierarchy ($m_1 < m_2 \ll m_3$). This gives a lower bound of $\sum m_\nu \gtrsim 0.1\,\mathrm{eV}$ and $\sum m_\nu \gtrsim 0.06\,\mathrm{eV}$ for the sum of the neutrino masses for the inverted and normal hierarchies respectively. An upper bound is given by $\sum m_\nu \lesssim 2.4\,\mathrm{eV}$, based on the KATRIN single $\beta$-decay experiment \citep{Aker2022}.

In addition to particle physics experiments, the sum of the neutrino masses can be constrained through cosmological observations. As neutrinos make up a fraction of the energy content of the Universe, given by \citep{Lesgourgues2006}
\begin{align}
    \Omega_{\nu} \approx \frac{\sum m_\nu}{93.14\,\mathrm{eV}\,h^2},
\end{align}
they affect the formation of structure. At early times, the massive neutrinos are relativistic, and free-stream out of overdense regions. This, in addition to alterations of the background evolution, like the time of matter-radiation equality, leads to a suppression of the matter power spectrum on scales smaller than the neutrino free-streaming length \citep{Lesgourgues2006}, 
\begin{align}
    \lambda_\mathrm{FS} = 7.7\frac{1+z}{\sqrt{\Omega_\Lambda + \Omega_m(1+z)^3}}\Bigg(\frac{1\,\mathrm{eV}}{\sum m_\nu}\Bigg)\,h^{-1}\mathrm{Mpc}.
\end{align}
Here, $\Omega_\mathrm{m} = \Omega_{\mathrm{cdm}} + \Omega_{\mathrm{b}} + \Omega_{\mathrm{\nu}}$ is the total energy density of matter and the other parameters are as explained before. The suppression of structure formation is observable and can help constrain the sum of the neutrino masses. A recent combination of various probes finds $\sum m_\nu \lesssim 0.09\,\mathrm{eV}$ at $95\%$ confidence \citep{Valentino2021} and one of the science goals of the \textit{Euclid} mission is to measure $\sum m_\nu$ to more than $0.03\,\mathrm{eV}$ precision through the use of weak gravitational lensing and galaxy clustering \citep{Laureijs2011}. 

Although cosmological observations can be used to obtain tighter upper bounds on the sum of the neutrino masses, it is important to take into account the dependence on the choice of a cosmological model. Hu-sawicki \fofr-modified gravity, as mentioned above, has the opposite effect of massive neutrinos on structure formation, and thus results in degenerate observables like the matter power spectrum, halo mass function (HMF), halo bias, and void-galaxy cross-correlation function \citep{Baldi2014,Mauland2023}.

\section{Method}
\label{sec:method}
In this section, we introduce the methods behind the simulations and machine learning codes used to create the emulator. We also detail the steps that need to be taken before applying the pipeline and the steps taken within the pipeline itself.

\subsection{Simulations}
\label{sec:cola}
The simulations in this paper were performed with the \colasolver{} implemented in the \texttt{FML} library\footnote{\url{https://github.com/HAWinther/FML/tree/master/FML/COLASolver}}. This is a fast and approximate particle-mesh (PM) \textit{N}-body code which employs the COLA method introduced by \cite{Tassev2013}. The \colasolver{} succeeds the \texttt{MG-PICOLA}\footnote{\url{https://github.com/HAWinther/MG-PICOLA-PUBLIC}} code \citep{Winther2017} and has various options for cosmologies and gravity models, including dynamical dark energy and \fofr{}-modified gravity. It also contains massive neutrinos, using a grid-based method as proposed in \cite{Brandebyge2009}, which is implemented and tested in \cite{Wright2017}. 

\subsubsection{The COLA method}
The COLA method \citep{Tassev2013} is based on the fact that structure formation on large scales is well described by Lagrangian perturbation theory (LPT). We can use this to our advantage and solve for the displacement, $\delta\bm{x}$, between a particle's LPT trajectory, $\bm{x}_\mathrm{LPT}$, and its full trajectory, $\bm{x}$. The geodesic equation for the particles is given by 
\begin{align}
    \frac{\mathrm{d}\bm{x}}{\mathrm{d}\tau} &= \bm{v},\\
    \frac{\mathrm{d}\bm{v}}{\mathrm{d}\tau} &= -\nabla\Phi,
\end{align}
which, when setting $\bm{x} = \delta\bm{x} + \bm{x}_\mathrm{LPT}$, becomes
\begin{align}
    \frac{\mathrm{d}\delta\bm{x}}{\mathrm{d}\tau} &= \delta\bm{v},\\
    \frac{\mathrm{d}\delta\bm{v}}{\mathrm{d}\tau} &= -\nabla\Phi - \frac{\mathrm{d}^2\bm{x}_{\rm LPT}}{\mathrm{d}\tau^2}.
\end{align}
The additional COLA force is easily computed from the displacement fields that are already calculated when creating the initial conditions. In this COLA frame (the frame co-moving with the LPT trajectories), the initial velocity of the particles is simply $\delta\bm{v} = 0$, and stays small on large scales during the evolution. This allows us to take much larger time steps than in usual {\it N}-body simulations, while still maintaining high accuracy on the largest scales, reducing the simulation time substantially. When increasing the number of timesteps, the method converges towards a full PM \textit{N}-body code. The COLA method has become an increasingly popular method for cheaply generating simulations and mock galaxy catalogues \citep{Tassev2015,Feng2016,Izard2016,Koda2016,Leclercq2020,Brando2023,Wright2023}.

\subsubsection{Screened modified gravity}
\label{sec:screened_mg}
The \colasolver{} we use already contains implementations of a wide range of modified gravity models, like $f(R)$ gravity, the symmetron, DGP, and Jordan-Brans-Dicke \citep{deFelice2010,Hinterbichler2011,Dvali2000,Joudaki2022}. A typical modified gravity model has a Poisson equation which in linear perturbation theory, and in Fourier space, reads \citep[see e.g.][]{Winther2017}
\begin{align}
    \Phi(k,z) = -\frac{3}{2k^2}\Omega_{m} a \delta_m(k,z) \frac{G_{\rm eff}(k,z)}{G}.
\end{align}
Here, $G_{\rm eff}(k,z)/G$ represents an effective Newtons constant, which might depend on both time and scale. For example, for the $f(R)$ model, we have
\begin{align}
    \frac{G_{\rm eff}(k,z)}{G} = 1 + \frac{1}{3} \frac{k^2}{k^2 + a^2m_{f(R)}^2},
\end{align}
where $m_{f(R)}^{-1}$ is the range of the fifth-force. This is exact on linear scales, but it does not include the important screening effect seen in many modified gravity models. To accurately take this into account, one must solve the non-linear partial differential equation (PDE) for the extra degree of freedom of the theory (e.g. the scalar field, $f_R$, for the case of $f(R)$ gravity). The \colasolver{} includes the possibility of doing exactly this, but it is quite time-consuming. Instead, we therefore rely on the method of \cite{WintherFerreira2015}. Here, the Poisson equation is taken to be
\begin{align}
    \Phi(k,z) = \Phi_N(k,z) - \frac{3}{2k^2}\Omega_m a\delta_m^{\rm eff}(k,z) \left(\frac{G_{\rm eff}(k,z)}{G}-1\right),
\end{align}
where the first term is standard Newtonian gravity and the second term is the contribution from the fifth force. The effective density, $\delta_m^{\rm eff}$, (in real space) is given by
\begin{align}
    \delta_m^{\rm eff}(x,z) = \delta_m(x,z) F(\Phi_N, \nabla\Phi_N, \nabla^2 \Phi_N, \ldots),
\end{align}
where the function $F$ estimates the screening. In this way, $F= 1$ corresponds to no screening. For different models, we can use spherical symmetry to compute the $F$ function. For example, for $f(R)$, we have
\begin{align}
    F = \min \left[1,\, \frac{3|f_{R0}|}{2|\Phi_N|}\left(\frac{\Omega_m + 4\Omega_\Lambda}{\Omega_m a^{-3} + 4\Omega_\Lambda}\right)^{n+1}\right],
\end{align}
which only depends on the local value of the standard Newtonian potential. This is easily (and cheaply) computed in the code using Fourier transforms, making the cost an order of magnitude lower than solving the full equation of motion.

In the \colasolver, different screening methods are already implemented for a wide range of models. The above approximation is accurate, but it is not perfect (depending on the model). Because of this, one should always compare the results to full {\it N}-body simulations, to assess the accuracy. If higher accuracy is needed, there is a possibility of improving it. One simple fix is to modify the screening method by introducing a fudge factor (or function), $\gamma(a)$, to scale $F$ with. Then, $\gamma(a)$ can be adjusted by comparing to exact simulations. This is done for $f(R)$ in \cite{WintherFerreira2015}, by fitting a constant factor to match a particular redshift. As the main purpose of this paper is to set up a general pipeline, and because emulators for the particular example model used here already exist \citep[e.g.][]{Ramachandra2021,Arnold2022,saez-casares2023}, we choose to not adjust $\gamma(a)$ and our screened simulations thus have $\gamma = 1$. The implication this has for the simulations we run is that it overestimates the screening, giving a conservative estimate for the actual boost with respect to $\Lambda$CDM.

\subsubsection{Massive neutrinos}

Massive neutrinos were for a long time considered beyond $\Lambda$CDM, at least from the perspective of \textit{N}-body simulations. This has changed over the last decade, and most simulations these days do include the effect of massive neutrinos.

In the \colasolver, massive neutrinos are treated as a field evolving according to linear theory, as proposed by \cite{Brandebyge2009}. After creating the CDM+baryon particles, we compute and store the initial density field, $\delta_{\rm cb}(k,z_{\rm ini})$, and evaluate
\begin{align}
    \delta_\nu(k,z) = \frac{T_\nu(k,z)}{T_\nu(k,z_{\rm ini})} \frac{T_\nu(k,z_{\rm init})}{T_{\rm cb}(k,z_{\rm init})} \delta_{\rm cb}(k,z_{\rm ini}),
\end{align}
where $T_{\rm cb}$ and $T_\nu$ are the CDM+baryon and neutrino transfer functions respectively. This is then added as a source to the Poisson equation (here for GR)
\begin{align}
    \Phi = -\frac{3}{2k^2}\Omega_{m} a \left[ (1-f_\nu)\delta_{\rm cdm} + f_\nu \delta_\nu \right],
\end{align}
where $f_\nu = \Omega_\nu / \Omega_m$. For more information about the neutrino implementation, see \cite{Wright2017}. The implementation of massive neutrinos used here is also included in a massive neutrino code comparison project \citep{Adamek2023}, where it shows percent level agreement in the power spectrum compared to more exact methods of including massive neutrinos.

\subsection{Machine learning}
\label{sec:machinelearning}
To create an emulator for the power spectrum boost, $B(k,z)$, we utilise \texttt{PyTorch-Lightening}\footnote{\url{https://www.pytorchlightning.ai/index.html}}, a lightweight wrapper for the Python \texttt{PyTorch} package \footnote{\url{https://github.com/pytorch/pytorch},\\\url{https://pytorch.org/}}. \texttt{PyTorch} is a machine-learning framework focusing on deep learning, and it provides the tools necessary to train neural networks with multiple layers. It requires our data as input, separated into three different categories: training, testing, and validation. The training data is used to train the neural network. This is the data that the neural network learns from. During the learning process, the neural network occasionally sees the validation data, as a means to help tune the model, but does not learn from it. Once the neural network is fully trained and the emulator is created, it can be evaluated against the test data to assess its performance. The architecture of the neural network training can be designed by the user by deciding the number of hidden dimensions, number of neurons, the batch size, and more \footnote{See the \texttt{PyTorch-Lightening} documentation for detailed instructions.}. The number of hidden dimensions governs how many layers there are between the input and output layers. Each of these layers has a given number of neurons, which perform computations on the training data before passing it to the next layer. The data is also commonly divided into smaller subsets, containing a set number of samples in each batch. This allows for a more efficient training process. 

For our neural network training, we assigned data to the training, validation, and test sets by drawing Latin hypercube samples \citep{McKay1979,Heitmann2006} for each data set, corresponding to a $80-10-10$ percent distribution. This ensured that each set had an even distribution of parameters in the available parameter set. We also tested both two and three hidden layers, in addition to varying the number of neurons in each layer, ranging from $8-512$ in different combinations. Finally, we tested the batch size, varying from $16-256$. From our example case of \fofr-modified gravity, we created three different emulators, two including unscreened \fofr{} gravity, one for the linear power spectrum boost and one for the non-linear, and one for the non-linear screened \fofr{} gravity power spectrum boost. They all had a batch size of $64$ and two hidden layers, but the two unscreened \fofr{} emulators used $16$ neurons in the first hidden layer and $8$ in the second, while the screened \fofr{} emulator had $32$ and $16$ neurons due to a more complicated shape for some of the curves. This can be further optimised and changed by the user based on the desired accuracy of the training process and will depend on the simulations and the parameters that are varied.

\subsection{Pipeline}
\label{sec_pipeline}
The full pipeline used for this work is made available to the public\footnote{\url{https://github.com/renmau/Sesame_pipeline}}, including instructions on how to use it. It can be applied as is for cosmologies with \fofr{} modified gravity and massive neutrinos, or extended to different cosmologies as wished. Here follows an outline of the steps necessary both to use the pipeline, and taken inside the pipeline itself:\\
First of all, the desired cosmological model, if not already included in the \colasolver{}, is implemented. Likewise, the model is implemented in a Boltzmann solver in order to obtain the initial conditions, or an already existing solver with the necessary cosmology can be used. Once this is done, the simulation setup is tested for the model in question to obtain the number of time steps, box size, grid resolution, and so on, that gives the desired convergence within the code itself. With the optimal setup obtained, the boosts dependence on cosmological parameters is tested in order to determine which parameters should be included when creating the emulator. When this is decided, the priors on the parameters is chosen, along with a fiducial cosmology and the number of desired samples to simulate for the neural network to work on. These are the steps that need to be taken outside of the pipeline. Once this is in order, Latin hypercube sampling is employed to sample the parameter space evenly. The way the pipeline is set up now, individual parameter samples are drawn for the training, testing, and validation sets so that they make up an $80-10-10$ percentile distribution of the total amount of samples. Alternatively, one can draw all the samples at once and then distribute the samples into data sets later. The desired amount of samples for the various cosmological parameters is written to file, together with the desired simulation settings of the \colasolver. The information in this file is then used to generate a bash script where new parameter files for the \colasolver{} are created. This script is then activated and the simulations are run for both the beyond-$\Lambda$CDM and \lcdm{} model for all the samples. This creates multiple outputs of the matter power spectrum at various redshifts in each case. The boost, $B(k,z) = P_{\text{beyond}-\Lambda \rm CDM}(k,z) / P_{\Lambda \rm CDM}(k, z)$, is then calculated for each parameter combination and redshift, and smoothed with a Savitzky-Golay filter \citep{Savitzky1964}. The smoothing is performed to reduce small fluctuations and thereby make the curves easier to estimate for the neural network. The parameter values, redshifts, $\log_{10}k$, and $B(k,z)$ are written into three separate files that go into the neural network learning. 80\% of the sample data goes into a training file, 10\% into a validation file, and the remaining 10\% into a test file, as mentioned above. These are then fed to the neural network, and the power spectrum ratio emulator is created. In this step, the architecture of the neural network must also be decided. This might take some trial and error, in order to obtain the desired accuracy.

At this point, the boost emulator for the desired cosmology has been created. To extract $P_{\text{beyond}-\Lambda \rm CDM}(k,z)$, we can now depend on already existing high-quality emulators for \lcdm{} \citep[e.g.][]{Angulo2021,knabenhans2021,Moran2023}. Ideally, the final step should be to run a high resolution \textit{N}-body simulation to determine the accuracy of the boost within the \colasolver. This would likely be the most expensive part to perform out of everything detailed above, but would give us an estimate of the simulation errors. Alternatively, if high-resolution simulations already exist for the cosmological model of interest, these can be used instead. In addition to this, we can get an estimate of the emulator errors by comparing the emulator performance to the test set. Both of these error estimates can then be baked into the covariance when using the emulator to fit to data, to ensure that all errors are included.

\section{Simulations}
\label{sec:simulations}
\begin{figure}
    \centering
    \includegraphics[width=0.5\textwidth]{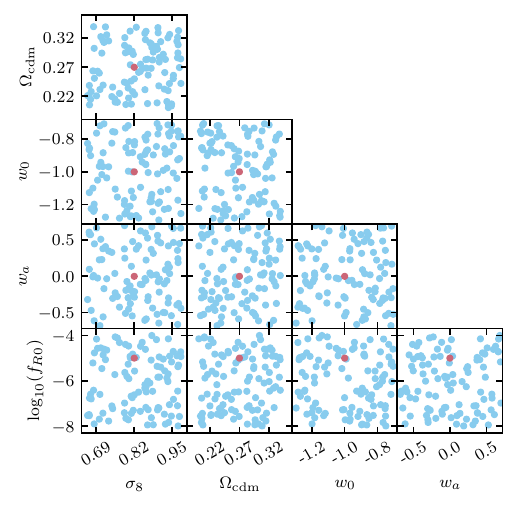}
    \caption{Sample distribution for 100 of the total 550 samples. The burgundy dot shows the fiducial cosmology parameter values, as given in Table~\ref{table:fiducial}.}
    \label{fig:sample}
\end{figure}

For the example case of the \fofr{} model, we performed two main sets of simulations with the \colasolver{} (Sect.~\ref{sec:cola}) to obtain the power spectrum boost. One setup had unscreened \fofr{} gravity, while the other included screening mechanisms. For these runs, we picked $550$ samples of five cosmological parameters, varied within the intervals
\begin{align}
    \sigma_8 &\in [0.66,0.98],\nonumber\\
    \Omega_\mathrm{cdm} &\in [0.20,0.34],\nonumber\\
    w_0 &\in [-1.3,-0.7],\\
    w_a &\in [-0.7,0.7],\nonumber\\
    \log_{10}f_{R0} &\in [-8.0,-4.0].\nonumber
\end{align}
These intervals, with the exception of $\log_{10}f_{R0}$, are based on the \texttt{EuclidEmulator2}, and are either the same ($w_0$, $w_a$) or slightly larger ($\Omega_\mathrm{cdm}$, $\sigma_8$) than the intervals used by \cite{knabenhans2021}\footnote{A convenience of using the \colasolver{} to perform the simulations is that it is fast, and therefore, extending the parameter intervals is computationally cheap compared to more accurate full \textit{N}-body simulations.}. A sample selection of 100 parameter samples can be seen in Fig.~\ref{fig:sample}.

\begin{figure}
    \centering
    \includegraphics[width=0.5\textwidth]{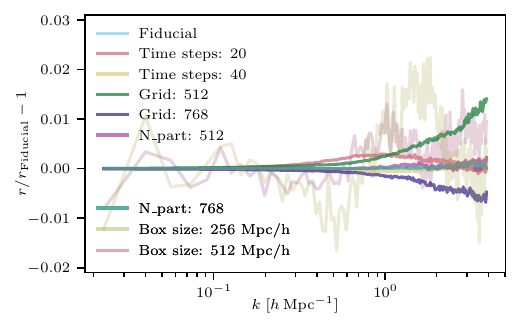}
    \caption{Convergence test for the \colasolver{} simulation setup for the screened boost between a \fofr{} gravity simulation with $|f_{R0}|=10^{-5}$ and $M_\nu=0.2\,\mathrm{eV}$, and a \lcdm{} simulation with massless neutrinos, at $z=0.0$. This ratio is denoted $r$, and is shown in comparison to a fiducial simulation setup. The fiducial setup is the same as the test setup in Table.~\ref{table:fiducial}, namely $N_\mathrm{time}=30$, $N_\mathrm{grid}=N_\mathrm{part}=640$, and $L_\mathrm{box}=350\,h^{-1}\mathrm{Mpc}$. Here, $N_\mathrm{time}$ denotes the number of time steps. The only parameter changed from the test setup to the final setup is $N_\mathrm{grid}$, which was increased to $N_\mathrm{grid}=786$ due to the resolution on non-linear scales.}
    \label{fig:sim_setup}
\end{figure}

\begin{figure}
    \centering
    \includegraphics[width=0.5\textwidth]{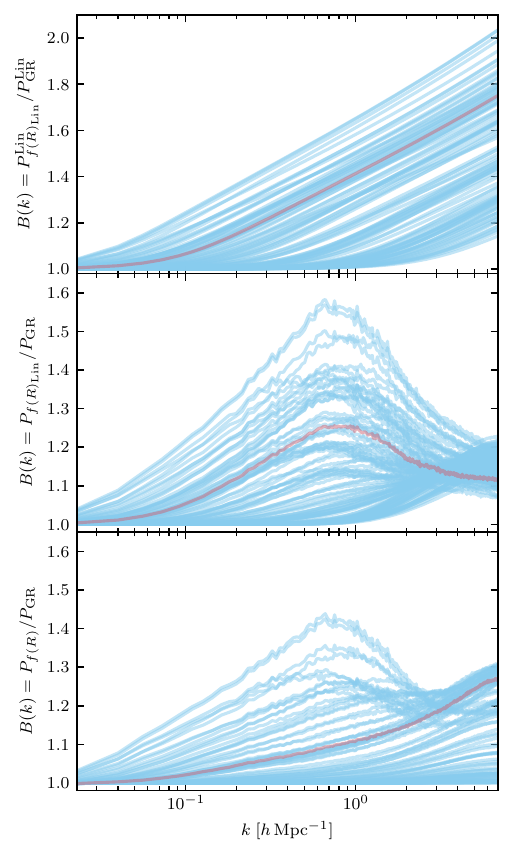}
    \caption{Matter power spectrum (CDM + baryons) boost for modified gravity and GR for the $100$ samples shown in Fig.~\ref{fig:sample} at $z=0.0$. The upper and middle panel shows the linear and non-linear boost for unscreened \fofr{} gravity. The lower panel shows the non-linear boost for screened \fofr{} gravity. The burgundy line displays the boost for the fiducial cosmology, as listed in Table \ref{table:fiducial}.}
    \label{fig:lin_ratio}
\end{figure}

The simulation setup and the fiducial cosmology are given in Table~\ref{table:fiducial}. In each case, we had $L_\mathrm{box}=350\,h^{-1}\mathrm{Mpc}$, $N_\mathrm{grid}=768$, and $N_\mathrm{part}=640$. The simulations were started at $z_\mathrm{ini}=30.0$ and used 30 timesteps up until $z=0.0$. Regarding the simulation setup we used, note that COLA simulations in general often use a large force-grid with $N_\mathrm{grid}=(2-3)N_\mathrm{part}$ \citep[see e.g.][]{Izard2016}. This is in order to have enough force-resolution to be able to create and resolve small halos - a crucial property if one is to create mock galaxy catalogues. The dark matter power-spectrum, on the other hand, is less sensitive to this, and we can therefore get away with using a smaller grid. When it comes to choosing the final simulation setup, it is important to always perform convergence tests of how the boost, $B$, changes with respect to the box size, the number of particles, the force resolution (the grid size), the number of time-steps, and other accuracy parameters like the initial redshift. This is essential to ensure that the result within COLA is converged. This is done for the setup used here, as seen in Fig.~\ref{fig:sim_setup}. Once this is done, the true accuracy can be assessed by comparing the COLA result to high-resolution \textit{N}-body simulations. The power spectrum boost for the 100 parameter samples mentioned above can be seen in Fig.~\ref{fig:lin_ratio} for three different scenarios; linear boost with unscreened \fofr{} gravity, non-linear boost with unscreened \fofr{} gravity, and non-linear boost with screened \fofr{} gravity.

\begin{table}
\caption{Fiducial values for the main simulations and for parameter variation tests run with a slightly different setup. If a parameter does not vary, this is its default value. The $M_\nu$ parameter refers to the sum of the neutrino masses and is given in eV. The main simulations have $L_\mathrm{box}=350\,h^{-1}\mathrm{Mpc}$, $N_\mathrm{grid}=768$, and $N_\mathrm{part}=640$. The parameter test runs have $L_\mathrm{box}=350\,h^{-1}\mathrm{Mpc}$ and $N_\mathrm{grid}=N_\mathrm{part}=640$.}           
\label{table:fiducial}      
\centering                          
\begin{tabular}{c@{\hskip 0.31in} c@{\hskip 0.31in} c}        
\hline\hline
\\[-0.3cm]
Parameter                 &Fiducial value     & Test value             \\    
\hline 
\\[-0.3cm]
    $A_s$                 & $2.1\times10^{-9}$ & $2.1\times10^{-9}$\\
    $\sigma_8$            & $0.82$             & $0.83$            \\
    $n_s$                 & $0.96$             & $0.96$            \\
    $h$                   & $0.67$             & $0.67$            \\
    $M_\nu$               & $0.058$            & $0.0$             \\
    $\Omega_\mathrm{cdm}$ & $0.27$             & $0.27$            \\
    $\Omega_\mathrm{b}$   & $0.049$            & $0.05$            \\
    $w_0$                 & $-1.0$             & $-1.0$            \\
    $w_a$                 & $0.0$              & $0.0$             \\
    $\log_{10}f_{R0}$     & $-5.0$             & $-5.0$            \\
\hline                                   
\end{tabular}
\end{table}

For every sample, \colasolver{} was run twice, once with \fofr{}-modified gravity and the selected value of $f_{R0}$, and once with regular GR. We ran our simulations for GR and $f(R)$ using the same initial conditions (i.e. we used the same value of $A_s$), which translates into
\begin{align}
\left(\sigma_8^{f(R)}\right)^2 = \int \frac{k^3}{2\pi^2} P_{\rm GR}(k,z=0) \left(\frac{D_{f(R)}(k,z)}{D_{\rm GR}(z)}\right)^2 \frac{{\rm d}k}{k},
\end{align}
where the growth factors, $D$, are normalised to unity in the early Universe. This ensures that the boost, $B$, is unity at early times, while today $\sigma_8^{f(R)}$ is slightly higher for our $f(R)$ simulations than our GR simulations. We used amplitude-fixed initial conditions \citep{Angulo2016,Villaescusa2018,Klypin2020} for our simulations to suppress the effects of cosmic variance.

\begin{figure}
    \centering
    \includegraphics[width=0.5\textwidth]{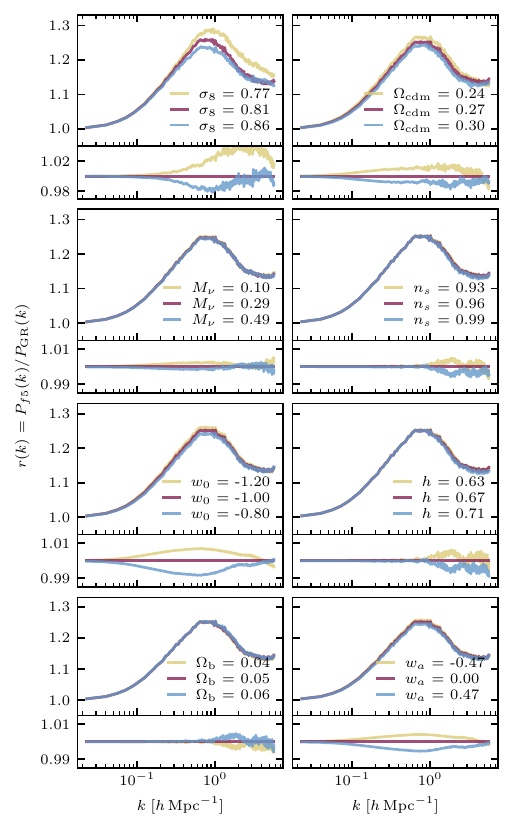}
    \caption{Boost for modified gravity with $f_{R0}=-10^{-5}$ and GR for different parameter variations. The ratios are shown in the larger panels for three different parameter values, while the narrower panels connected to each large panel show the corresponding ratios of ratios for the three different parameter values, with the middle value as the baseline. The parameters are varied while holding the rest constant, and the fiducial test cosmology is given in Table~\ref{table:fiducial}. When varying $\Omega_\mathrm{cdm}$, $\Omega_\mathrm{b}$ is kept constant, meaning that $\Omega_\mathrm{m}$ varies accordingly. When $\Omega_\mathrm{b}$ is varied, $\Omega_\mathrm{cdm}$ also varies so that $\Omega_\mathrm{m}$ is kept constant at a value of $0.32$. There is no screening invoked for the \fofr{} simulations. The $M_\nu$ parameter refers to the sum of the neutrino masses and is given in eV. Note the different axes for the first row narrow panels compared to the rest.}
    \label{fig:linf5_fixeds8_varys8}
\end{figure}

In order to know which cosmological parameter to include as variables in the emulator training, we performed some test simulations. Figure~\ref{fig:linf5_fixeds8_varys8} displays the non-linear boost for the unscreened \fofr{} case, with $|f_{R0}|=10^{-5}$, when different parameters are allowed to vary. From this, it is clear that $\sigma_8$, $\Omega_\mathrm{cdm}$, $w_0$, and possibly $w_a$ are the most influential parameters on the power spectrum ratio. Because of this, $\sigma_8$, $\Omega_\mathrm{cdm}$, $w_0$, $w_a$, and $\log_{10}f_{R0}$, in addition to $z$ and $k$, were chosen as the parameters to vary when producing the data used to train the neural network when creating the boost emulator. Tests performed where $\sigma_8$ was not kept fixed for the \fofr{} and GR initial conditions showed a larger variation for all the parameters in general. However, fixing $\sigma_8$ showed that some of this effect was due to the difference in clustering. We also performed tests with screened \fofr{} gravity and a different value for $f_{R0}$. These tests also pointed toward the same parameter choices. 

\section{Results}
\label{sec:results}
\begin{figure}
    \centering
    \includegraphics[width=0.5\textwidth]
    {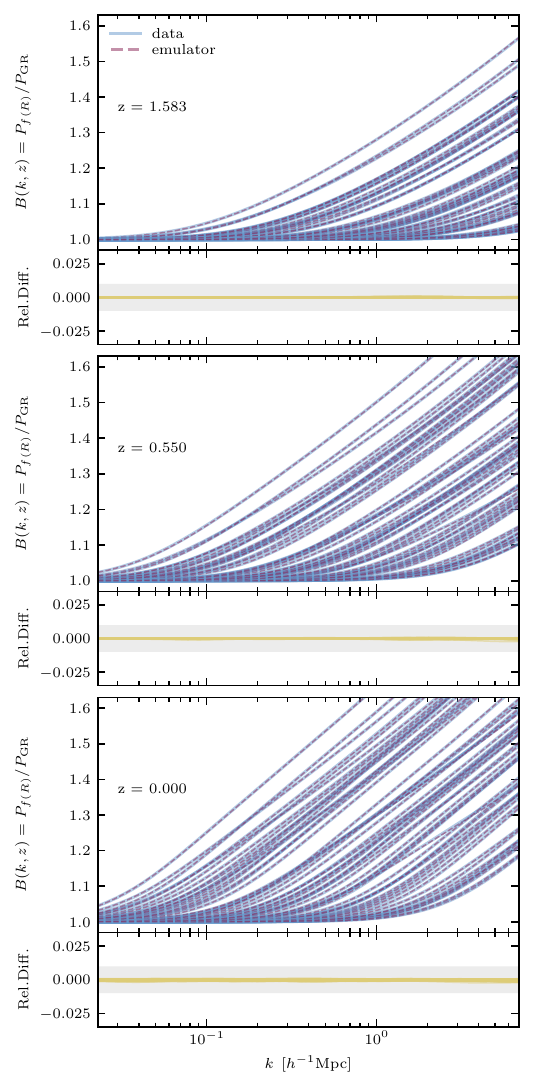}
    \caption{Emulator performance compared to the test data sets for various redshifts for the linear boost with unscreened \fofr{} gravity. The emulator results, along with the simulations, are given in the larger panels, while the narrower panels display the corresponding relative difference, given by $B_\mathrm{emulator}/B_\mathrm{simulation} -1$. The grayed-out area in the same panel shows $\pm1\%$, and the Nyquist frequency of the simulations is $k\approx5.7\,h\,\mathrm{Mpc}^{-1}$.}
    \label{fig:lin_pk_lin_fofr}
\end{figure}

\begin{figure}
    \centering
    \includegraphics[width=0.5\textwidth]
    {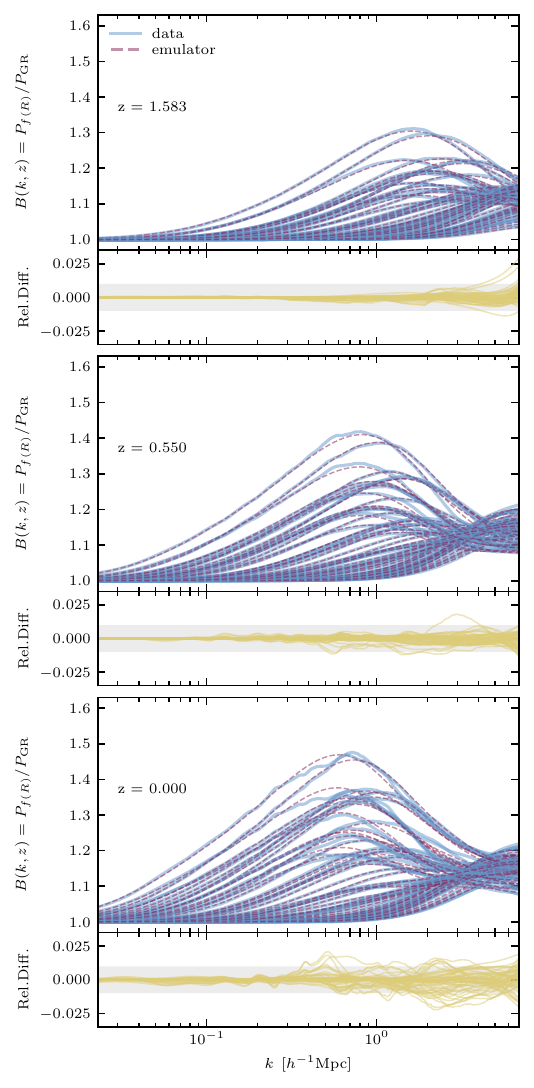}
    \caption{Emulator performance compared to the test data sets for various redshifts for the non-linear boost with unscreened \fofr{} gravity. Figure setup as explained for Fig.~\ref{fig:lin_pk_lin_fofr}.}
    \label{fig:nonlin_pk_lin_fofr}
\end{figure}

\begin{figure}
    \centering
    \includegraphics[width=0.5\textwidth]
    {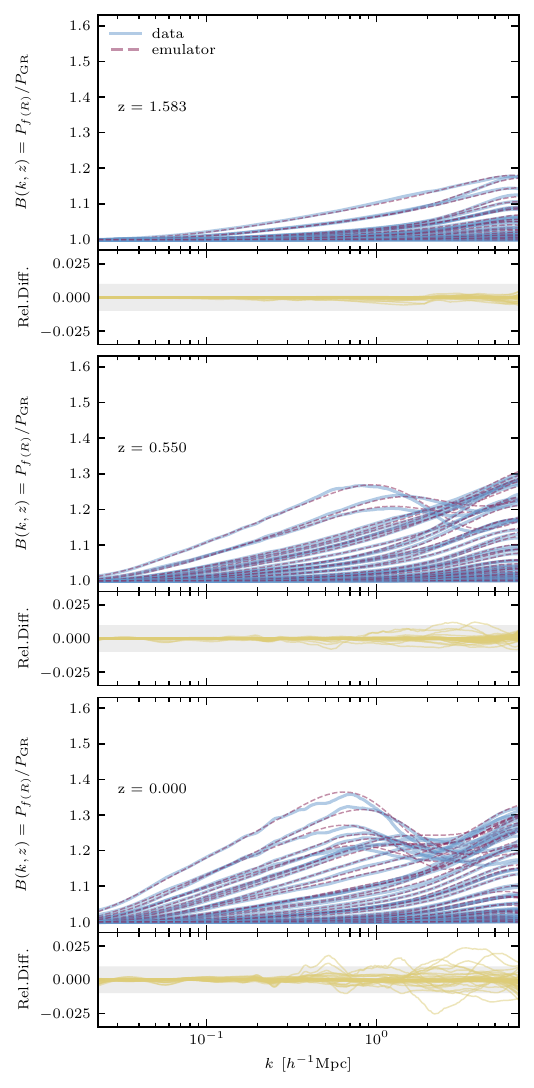}
    \caption{Emulator performance compared to the test data sets for various redshifts for the non-linear boost with screened \fofr{} gravity. Figure setup as explained for Fig.~\ref{fig:lin_pk_lin_fofr}.}
    \label{fig:nonlin_pk_nonlin_fofr}
\end{figure}

In this section, we present the results for our example boost emulator with \fofr-modified gravity for three different cases: the linear power spectrum boost with unscreened \fofr{} gravity, the non-linear power spectrum boost with unscreened \fofr{} gravity, and the non-linear power spectrum boost with screened \fofr{} gravity. The emulator results compared to the test data sets can be seen in Figs.~\ref{fig:lin_pk_lin_fofr}, \ref{fig:nonlin_pk_lin_fofr}, and \ref{fig:nonlin_pk_nonlin_fofr} respectively, for three different redshifts.

\begin{figure}
    \centering
    \includegraphics[width=0.5\textwidth]
    {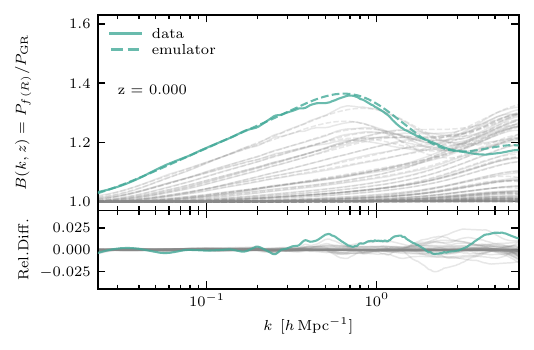}
    \caption{Lower panel of Fig.~\ref{fig:nonlin_pk_nonlin_fofr} with one of the largest outliers highlighted in green. The rest of the results are as before, but muted with a gray colour. The parameter sample values of the outlier can be found in Table.~\ref{table:outlier}.}
    \label{fig:nonlin_pk_nonlin_fofr_outlier}
\end{figure}

In general, we see that the fully linear case has better agreement between the emulator and test data for all redshifts, compared to the non-linear cases. This is most likely due to the simplicity of the boost curve, making it easier for the neural network to predict. The same effect is also seen for higher redshifts in all three cases, where the curves flatten out and become easier for the learning processes to capture accurately. Overall, the fully linear boost emulator agrees with the \colasolver{} simulations to below one percent accuracy on all scales and all redshifts (Fig.~\ref{fig:lin_pk_lin_fofr}). For the non-linear unscreened \fofr{} case (Fig.~\ref{fig:nonlin_pk_lin_fofr}), we have agreement to below or around $1\%$ at all scales for redshifts $z=1.58$ and $z=0.55$, with only a few exceptions showing slightly larger discrepancies at small scales. For $z=0.00$, we have below $1\%$ agreement up to $k\sim0.3\,h\,\mathrm{Mpc}^{-1}$, and between $2-2.5\%$ agreement otherwise. For the non-linear boost in the screened \fofr{} gravity case (Fig.~\ref{fig:nonlin_pk_nonlin_fofr}), we have below or around one percent accuracy at all scales for $z=1.58$ and $z=0.55$. When we reach redshift zero there are a few outliers, resulting in some differences around $2.5\%$, although the bulk of the set stays below $1\%$. Still, for the non-linear boost emulator, both in the case of screened and unscreened \fofr{} gravity, it is clear, when compared to Fig.~\ref{fig:lin_ratio}, that the curves with the largest discrepancy between predictions and simulations can differ quite a lot from the fiducial expectation. This is not unexpected, as the training set for the neural network contains fewer samples with parameter values that lie close to the edges of the allowed intervals, therefore making the predictions less robust for periphery samples. An example of this is shown in Fig.~\ref{fig:nonlin_pk_nonlin_fofr_outlier}, where an outlier is highlighted. The corresponding parameter sample, compared to the fiducial values, is given in Table~\ref{table:outlier}. The error can be further approved by adjusting the neural network architecture, but this depends on the features in the curve and the parameters included in the training process, and must therefore be adjusted individually for anyone interested in applying the pipeline. It should also be mentioned that for the screened \fofr{} gravity emulator (Fig.~\ref{fig:nonlin_pk_nonlin_fofr}), there could be some overfitting for the simplest curves, due to the relatively complex architecture containing $32$ and $16$ neurons in the two hidden layers. We found that this was necessary in order to catch the shape of the more complex curves, like the one highlighted in Fig.~\ref{fig:nonlin_pk_nonlin_fofr_outlier}. This could possibly be remedied by supplying the neural network with smoother data curves. 

\begin{figure}
    \centering
    \includegraphics[width=0.5\textwidth]{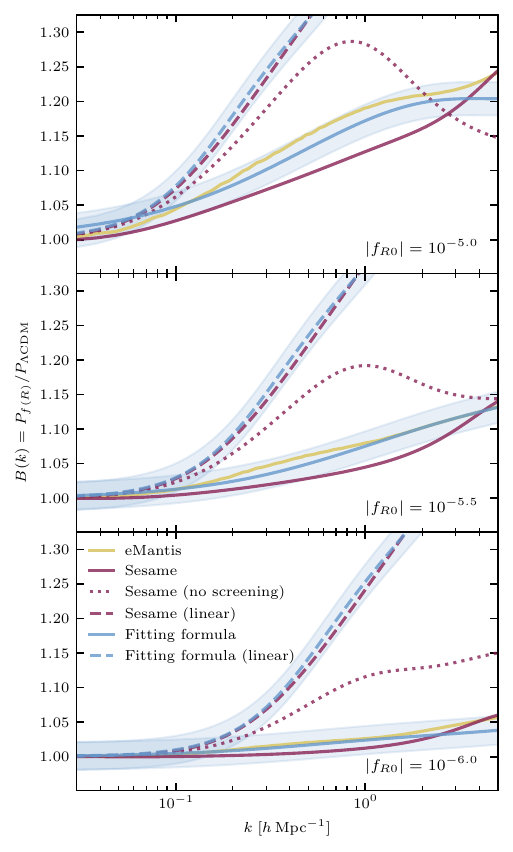}
    \caption{Comparison of our emulators versus emulators in the literature for $z=0$. The green line shows e-Mantis \citep{saez-casares2023} - an emulator based on high-resolution simulations. The red lines show the fitting formula of \cite{Winther2019}, which is based on high-resolution simulations for a fixed cosmology. The shaded bands highlight $\pm 2\%$ about the fitting formula prediction.}
    \label{fig:comparison}
\end{figure}

Finally, in Fig.~\ref{fig:comparison}, we make a comparison with a small selection of other emulators in the literature for one set of cosmological parameters (corresponding to the parameters used to make the fitting formula of \cite{Winther2019}). Overall, we find a good agreement. The linear emulator is within $\sim 1\%$ of the linear fitting formula, while the non-linear emulator is within 2-3 \% of e-Mantis \citep{saez-casares2023} and the fitting formula. Our no-screening prediction naturally falls between the linear and the non-linear predictions. To our knowledge, there are no other emulators available to compare this to, but for $|f_{R0}| = 10^{-4}$, where screening is not very active, it agrees to a few percent with e-Mantis and the fitting formula. The non-linear emulator we have created here overestimates the screening and gives a conservative estimate for the boost. By properly calibrating the screening efficiency, $\gamma(a)$, in the approximate screening method we use we could improve on this result. As the emulator we created here is mainly an example of the emulator pipeline, we have foregone this step. 

\begin{table}
\caption{Parameters for one of the samples with the largest discrepancy between simulated and emulated boost for the screened \fofr{} gravity case, compared to the fiducial values. The parameters $\Omega_\mathrm{cdm}$, $w_a$, and $\log_{10}f_{R0}$ all have values close to the interval boundaries of the emulator training data.}           
\label{table:outlier}      
\centering                          
\begin{tabular}{c@{\hskip 0.31in} c@{\hskip 0.31in} c}        
\hline\hline
\\[-0.3cm]
Parameter                 &Fiducial sample     & Outlier sample    \\    
\hline 
\\[-0.3cm]
    $\sigma_8$            & $0.82$             & $0.87$            \\
    $\Omega_\mathrm{cdm}$ & $0.27$             & $0.21$            \\
    $w_0$                 & $-1.0$             & $-1.1$            \\
    $w_a$                 & $0.00$              & $-0.46$           \\
    $\log_{10}f_{R0}$     & $-5.00$             & $-4.04$            \\
\hline                                   
\end{tabular}
\end{table}

\section{Conclusions}
\label{sec:conclusions}
Emulators for various global clustering statistics are memory and time-saving. However, creating them often requires a lot of resources through the use of large \textit{N}-body simulation suites. Because of this, the construction of accurate emulators usually depends on the use of supercomputers. In this paper, we present a full pipeline, \sesame, for creating emulators for the matter power spectrum boost, $B(k,z) = P_{\text{beyond}-\Lambda \rm CDM}(k,z) / P_{\Lambda \rm CDM}(k, z)$, for beyond-\lcdm{} models, without the need for large computing resources. The pipeline employs the fast and approximate COLA method \citep{Tassev2013, Wright2017,Winther2017} to perform the simulations, simulating both the beyond-\lcdm{} and \lcdm{} model. This allows us to extract the boost up to higher $k$-values, due to some of the internal code artifacts canceling, as demonstrated in for example \cite{Adamek2023}. The simulation data is then used to train a neural network, through the \texttt{PyTorch Lightening} deep learning module, resulting in a boost emulator. At this point, we rely on existing \lcdm{} emulators to extract $P_{\text{beyond}-\Lambda \rm CDM}(k,z)$.

Using the pipeline will consist of the following steps:
\begin{itemize}
    \item Implement the model or parametrisation you want to emulate in the \colasolver{}. This most commonly consists of implementing the \textit{Hubble} function and how to compute the gravitational potential. Here, already implemented models can be used as examples. For most models, this will be a minor task.
    \item Pick the simulation setup and do a convergence test to ensure that the setup is converged within the code itself.
    \item Pick which cosmological parameters you are interested in varying. For this, it is useful to study how the boost, $B$, changes when varying individual cosmological parameters, and select the ones that have a significant impact. From our experience, looking at different modified gravity models that deviate from $\Lambda$CDM only close to today, as long as the power ($\sigma_8$) is kept the same (depending on the model), either in the initial conditions or at $z=0$, it is often $\sigma_8$ (or $A_s$) and $\Omega_m$ that are the most relevant.    
    \item Pick the priors of the parameters you want to vary and the number of samples you want to include, and use this to generate the Latin hypercube samples (script provided in the pipeline).
    \item Generate all the input for \colasolver{}, meaning the input files and the necessary power spectra, by running a Boltzmann solver (script provided in the pipeline for \texttt{CLASS} \citep{Lesgourgues2011, Blas2011}).
    \item Run the simulations to produce all the data files containing the power spectra needed to compute the boosts (script provided in the pipeline).
    \item Gather all the data and make the files needed for the emulator (script provided in the pipeline). 
    \item Determine the neural network architecture (often trial and error) and run the training to produce the emulator (script provided in the pipeline).
    \item Check the accuracy of the emulator and redo the previous step if needed until you have something acceptable (script to compare the emulator with data provided in the pipeline).
    \item Estimate the errors. The emulation error can be obtained from the training set and the error of the simulations themselves can be estimated by running a set of high-resolution \textit{N}-body simulations or by using already existing simulations.
\end{itemize}

As an example of using this pipeline, we created three emulators for \fofr-modified gravity, including massive neutrinos. The three emulators estimate the boost in the cases of linear and non-linear boost for unscreened \fofr{} gravity, and the non-linear boost for screened \fofr{} gravity. The first two of these have not been made before, while for the last case there already exists several emulators \citep[e.g.][]{Ramachandra2021,Arnold2022,saez-casares2023}. 

The fully linear emulator has below-percent accuracy compared to ground truth, while the non-linear boost emulators have around $1-2\%$ accuracy at redshift zero compared to our simulations and $3-4\%$ accuracy compared to high-resolution simulations. We stress that the emulator with screening that we created here is mainly an example of using the pipeline and is not meant to rival high-quality emulators such as e-Mantis, which is based on high-resolution simulations solving the full $f(R)$ field equation. If this was to be the case, the screening efficiency in the approximate screening method we use for $f(R)$ would have had to be calibrated by comparing to full simulations to enhance the accuracy of the COLA approach. As this is not done here, the non-linear emulator we have created overestimates the screening and gives a lower estimate for the boost. When using emulators to fit data, both the error between the emulator and COLA simulations and that of the COLA method compared to \textit{N}-body simulations must be taken into account.

With the paper, we provide the full pipeline\footnote{\url{https://github.com/renmau/Sesame_pipeline}}. \sesame{} can then be used by anyone to create emulators for their desired beyond-\lcdm{} model, either by employing one of the models already incorporated in \colasolver{} code, or by implementing the desired model and then applying the pipeline.

\begin{acknowledgements}
We would like to thank the Research Council of Norway for their support.
\end{acknowledgements}

\bibliographystyle{aa} 
\bibliography{main} 

\begin{thebibliography}{96}
\expandafter\ifx\csname natexlab\endcsname\relax\def\natexlab#1{#1}\fi

\bibitem[{Adamek {et~al.}(2017)Adamek, Durrer, \& Kunz}]{Adamek2017}
Adamek, J., Durrer, R., \& Kunz, M. 2017, Journal of Cosmology and Astroparticle Physics, 2017, 004

\bibitem[{Amendola(2000)}]{Amendola2000}
Amendola, L. 2000, Physical Review D, 62, 043511

\bibitem[{Angulo \& Pontzen(2016)}]{Angulo2016}
Angulo, R.~E. \& Pontzen, A. 2016, Monthly Notices of the Royal Astronomical Society: Letters, 462, L1

\bibitem[{Angulo {et~al.}(2021)Angulo, Zennaro, Contreras, Aric{\`{o}}, Pellejero-Iba{\~{n}}ez, \& St{\"{u}}cker}]{Angulo2021}
Angulo, R.~E., Zennaro, M., Contreras, S., {et~al.} 2021, Monthly Notices of the Royal Astronomical Society, 507, 5869

\bibitem[{Arnold {et~al.}(2022)Arnold, Li, Giblin, Harnois-D{\'{e}}raps, \& Cai}]{Arnold2022}
Arnold, C., Li, B., Giblin, B., Harnois-D{\'{e}}raps, J., \& Cai, Y.~C. 2022, Monthly Notices of the Royal Astronomical Society, 515, 4161

\bibitem[{Baldi {et~al.}(2014)Baldi, Villaescusa-Navarro, Viel, Puchwein, Springel, \& Moscardini}]{Baldi2014}
Baldi, M., Villaescusa-Navarro, F., Viel, M., {et~al.} 2014, Monthly Notices of the Royal Astronomical Society, 440, 75

\bibitem[{Blas {et~al.}(2011)Blas, Lesgourgues, \& Tram}]{Blas2011}
Blas, D., Lesgourgues, J., \& Tram, T. 2011, Journal of Cosmology and Astroparticle Physics, 2011, 034

\bibitem[{Bose {et~al.}(2020)Bose, Cataneo, Tr{\"{o}}ster, Xia, Heymans, \& Lombriser}]{Bose2020}
Bose, B., Cataneo, M., Tr{\"{o}}ster, T., {et~al.} 2020, Monthly Notices of the Royal Astronomical Society, 498, 4650

\bibitem[{Bose {et~al.}(2023)Bose, Tsedrik, Kennedy, Lombriser, Pourtsidou, \& Taylor}]{Bose2023}
Bose, B., Tsedrik, M., Kennedy, J., {et~al.} 2023, Monthly Notices of the Royal Astronomical Society, 519, 4780

\bibitem[{Bose {et~al.}(2021)Bose, Wright, Cataneo, Pourtsidou, Giocoli, Lombriser, Mccarthy, Baldi, Pfeifer, \& Xia.}]{Bose2021}
Bose, B., Wright, B.~S., Cataneo, M., {et~al.} 2021, Monthly Notices of the Royal Astronomical Society, 508, 2479

\bibitem[{Brandbyge \& Hannestad(2009)}]{Brandebyge2009}
Brandbyge, J. \& Hannestad, S. 2009, Journal of Cosmology and Astroparticle Physics, 2009, 002

\bibitem[{Brando {et~al.}(2022)Brando, Fiorini, Koyama, \& Winther}]{Brando2022}
Brando, G., Fiorini, B., Koyama, K., \& Winther, H.~A. 2022, Journal of Cosmology and Astroparticle Physics, 2022, 051

\bibitem[{Brando {et~al.}(2023)Brando, Koyama, \& Winther}]{Brando2023}
Brando, G., Koyama, K., \& Winther, H.~A. 2023, Journal of Cosmology and Astroparticle Physics, 2023, 045

\bibitem[{Brax {et~al.}(2008)Brax, Van De~Bruck, Davis, \& Shaw}]{Brax2008}
Brax, P., Van De~Bruck, C., Davis, A.~C., \& Shaw, D.~J. 2008, Physical Review D - Particles, Fields, Gravitation and Cosmology, 78, 104021

\bibitem[{Buchdahl(1970)}]{Buchdahl1970}
Buchdahl, H.~A. 1970, Monthly Notices of the Royal Astronomical Society, 150, 1

\bibitem[{Bull {et~al.}(2016)Bull, Akrami, Adamek, Baker, Bellini, Beltr{\'{a}}n~Jim{\'{e}}nez, Bentivegna, Camera, Clesse, Davis, Di~Dio, Enander, Heavens, Heisenberg, Hu, Llinares, Maartens, M{\"{o}}rtsell, Nadathur, Noller, Pasechnik, Pawlowski, Pereira, Quartin, Ricciardone, Riemer-S{\o}rensen, Rinaldi, Sakstein, Saltas, Salzano, Sawicki, Solomon, Spolyar, Starkman, Steer, Tereno, Verde, Villaescusa-Navarro, von Strauss, \& Winther}]{Bull2016}
Bull, P., Akrami, Y., Adamek, J., {et~al.} 2016, Physics of the Dark Universe, 12, 56

\bibitem[{Cataneo {et~al.}(2015)Cataneo, Rapetti, Schmidt, Mantz, Allen, Applegate, Kelly, Von Der~Linden, \& Morris}]{Cataneo2015}
Cataneo, M., Rapetti, D., Schmidt, F., {et~al.} 2015, Physical Review D - Particles, Fields, Gravitation and Cosmology, 92, 044009

\bibitem[{Chabanier {et~al.}(2019)Chabanier, Millea, \& Palanque-Delabrouille}]{Chabanier2019}
Chabanier, S., Millea, M., \& Palanque-Delabrouille, N. 2019, Monthly Notices of the Royal Astronomical Society, 489, 2247

\bibitem[{Chevallier \& Polarski(2001)}]{Chevallier2001}
Chevallier, M. \& Polarski, D. 2001, International Journal of Modern Physics D, 10, 213

\bibitem[{Clifton {et~al.}(2012)Clifton, Ferreira, Padilla, \& Skordis}]{Clifton2012}
Clifton, T., Ferreira, P.~G., Padilla, A., \& Skordis, C. 2012, Physics Reports, 513, 1

\bibitem[{Dakin {et~al.}(2019)Dakin, Brandbyge, Hannestad, Haugb{\o}lle, \& Tram}]{Dakin2019}
Dakin, J., Brandbyge, J., Hannestad, S., Haugb{\o}lle, T., \& Tram, T. 2019, Journal of Cosmology and Astroparticle Physics, 2019, 052

\bibitem[{de~Felice \& Tsujikawa(2010)}]{deFelice2010}
de~Felice, A. \& Tsujikawa, S. 2010, Living Reviews in Relativity, 13, 1

\bibitem[{{DES Collaboration} {et~al.}(2021){DES Collaboration}, Abbott, Adam{\'{o}}w, Aguena, Allam, Amon, Annis, Avila, Bacon, Banerji, Bechtol, Becker, Bernstein, Bertin, Bhargava, Bridle, Brooks, Burke, Rosell, Kind, Carretero, Castander, Cawthon, Chang, Choi, Conselice, Costanzi, Crocce, Costa, Davis, Vicente, DeRose, Desai, Diehl, Dietrich, Drlica-Wagner, Eckert, Elvin-Poole, Everett, Evrard, Ferrero, Fert{\'{e}}, Flaugher, Fosalba, Friedel, Frieman, Garc{\'{i}}a-Bellido, Gaztanaga, Gelman, Gerdes, Giannantonio, Gill, Gruen, Gruendl, Gschwend, Gutierrez, Hartley, Hinton, Hollowood, Honscheid, Huterer, James, Jeltema, Johnson, Kent, Kron, Kuehn, Kuropatkin, Lahav, Li, Lidman, Lin, MacCrann, Maia, Manning, Maloney, March, Marshall, Martini, Melchior, Menanteau, Miquel, Morgan, Myles, Neilsen, Ogando, Palmese, Paz-Chinch{\'{o}}n, Petravick, Pieres, Plazas, Pond, Rodriguez-Monroy, Romer, Roodman, Rykoff, Sako, Sanchez, Santiago, Scarpine, Serrano, Sevilla-Noarbe, Smith, Smith, Soares-Santos, Suchyta,
  Swanson, Tarle, Thomas, To, Tremblay, Troxel, Tucker, Turner, Varga, Walker, Wechsler, Weller, Wester, Wilkinson, Yanny, Zhang, Nikutta, Fitzpatrick, Jacques, Scott, Olsen, Huang, Herrera, Juneau, Nidever, Weaver, Adean, Correia, Freitas, Freitas, Singulani, Vila-Verde, \& Server)}]{DES2021}
{DES Collaboration}, Abbott, T. M.~C., Adam{\'{o}}w, M., {et~al.} 2021, The Astrophysical Journal Supplement Series, 255, 20

\bibitem[{{DESI Collaboration} {et~al.}(2016){DESI Collaboration}, Aghamousa, Aguilar, Ahlen, Alam, Allen, Allende~Prieto, Annis, Bailey, Balland, Ballester, Baltay, Beaufore, Bebek, Beers, Bell, Bernal, Besuner, Beutler, Blake, Bleuler, Blomqvist, Blum, Bolton, Briceno, Brooks, Brownstein, Buckley-Geer, Burden, Burtin, Busca, Cahn, Cai, Cardiel-Sas, Carlberg, Carton, Casas, Castander, Cervantes-Cota, Claybaugh, Close, Coker, Cole, Comparat, Cooper, Cousinou, Crocce, Cuby, Cunningham, Davis, Dawson, de~la Macorra, De~Vicente, Delubac, Derwent, Dey, Dhungana, Ding, Doel, Duan, Ealet, Edelstein, Eftekharzadeh, Eisenstein, Elliott, Escoffier, Evatt, Fagrelius, Fan, Fanning, Farahi, Farihi, Favole, Feng, Fernandez, Findlay, Finkbeiner, Fitzpatrick, Flaugher, Flender, Font-Ribera, Forero-Romero, Fosalba, Frenk, Fumagalli, Gaensicke, Gallo, Garcia-Bellido, Gaztanaga, Pietro Gentile~Fusillo, Gerard, Gershkovich, Giannantonio, Gillet, Gonzalez-de Rivera, Gonzalez-Perez, Gott, Graur, Gutierrez, Guy, Habib, Heetderks,
  Heetderks, Heitmann, Hellwing, Herrera, Ho, Holland, Honscheid, Huff, Hutchinson, Huterer, Hwang, Illa~Laguna, Ishikawa, Jacobs, Jeffrey, Jelinsky, Jennings, Jiang, Jimenez, Johnson, Joyce, Jullo, Juneau, Kama, Karcher, Karkar, Kehoe, Kennamer, Kent, Kilbinger, Kim, Kirkby, Kisner, Kitanidis, Kneib, Koposov, Kovacs, Koyama, Kremin, Kron, Kronig, Kueter-Young, Lacey, Lafever, Lahav, Lambert, Lampton, Landriau, Lang, Lauer, Le~Goff, Le~Guillou, Le~Van~Suu, Lee, Lee, Leitner, Lesser, Levi, L'Huillier, Li, Liang, Lin, Linder, Loebman, Luki{\'{c}}, Ma, MacCrann, Magneville, Makarem, Manera, Manser, Marshall, Martini, Massey, Matheson, McCauley, McDonald, McGreer, Meisner, Metcalfe, Miller, Miquel, Moustakas, Myers, Naik, Newman, Nichol, Nicola, Nicolati~da Costa, Nie, Niz, Norberg, Nord, Norman, Nugent, O'Brien, Oh, Olsen, Padilla, Padmanabhan, Padmanabhan, Palanque-Delabrouille, Palmese, Pappalardo, P{\^{a}}ris, Park, Patej, Peacock, Peiris, Peng, Percival, Perruchot, Pieri, Pogge, Pollack, Poppett, Prada,
  Prakash, Probst, Rabinowitz, Raichoor, Ree, Refregier, Regal, Reid, Reil, Rezaie, Rockosi, Roe, Ronayette, Roodman, Ross, Ross, Rossi, Rozo, Ruhlmann-Kleider, Rykoff, Sabiu, Samushia, Sanchez, Sanchez, Schlegel, Schneider, Schubnell, Secroun, Seljak, Seo, Serrano, Shafieloo, Shan, Sharples, Sholl, Shourt, Silber, Silva, Sirk, Slosar, Smith, Smoot, Som, Song, Sprayberry, Staten, Stefanik, Tarle, Sien~Tie, Tinker, Tojeiro, Valdes, Valenzuela, Valluri, Vargas-Magana, Verde, Walker, Wang, Wang, Weaver, Weaverdyck, Wechsler, Weinberg, White, Yang, Yeche, Zhang, Zhao, Zheng, Zhou, Zhou, Zhu, Zou, Zu, Collaboration, Aghamousa, Aguilar, Ahlen, Alam, Allen, Allende~Prieto, Annis, Bailey, Balland, Ballester, Baltay, Beaufore, Bebek, Beers, Bell, Bernal, Besuner, Beutler, Blake, Bleuler, Blomqvist, Blum, Bolton, Briceno, Brooks, Brownstein, Buckley-Geer, Burden, Burtin, Busca, Cahn, Cai, Cardiel-Sas, Carlberg, Carton, Casas, Castander, Cervantes-Cota, Claybaugh, Close, Coker, Cole, Comparat, Cooper, Cousinou, Crocce,
  Cuby, Cunningham, Davis, Dawson, de~la Macorra, De~Vicente, Delubac, Derwent, Dey, Dhungana, Ding, Doel, Duan, Ealet, Edelstein, Eftekharzadeh, Eisenstein, Elliott, Escoffier, Evatt, Fagrelius, Fan, Fanning, Farahi, Farihi, Favole, Feng, Fernandez, Findlay, Finkbeiner, Fitzpatrick, Flaugher, Flender, Font-Ribera, Forero-Romero, Fosalba, Frenk, Fumagalli, Gaensicke, Gallo, Garcia-Bellido, Gaztanaga, Pietro Gentile~Fusillo, Gerard, Gershkovich, Giannantonio, Gillet, Gonzalez-de Rivera, Gonzalez-Perez, Gott, Graur, Gutierrez, Guy, Habib, Heetderks, Heetderks, Heitmann, Hellwing, Herrera, Ho, Holland, Honscheid, Huff, Hutchinson, Huterer, Hwang, Illa~Laguna, Ishikawa, Jacobs, Jeffrey, Jelinsky, Jennings, Jiang, Jimenez, Johnson, Joyce, Jullo, Juneau, Kama, Karcher, Karkar, Kehoe, Kennamer, Kent, Kilbinger, Kim, Kirkby, Kisner, Kitanidis, Kneib, Koposov, Kovacs, Koyama, Kremin, Kron, Kronig, Kueter-Young, Lacey, Lafever, Lahav, Lambert, Lampton, Landriau, Lang, Lauer, Le~Goff, Le~Guillou, Le~Van~Suu, Lee, Lee,
  Leitner, Lesser, Levi, L'Huillier, Li, Liang, Lin, Linder, Loebman, Luki{\'{c}}, Ma, MacCrann, Magneville, Makarem, Manera, Manser, Marshall, Martini, Massey, Matheson, McCauley, McDonald, McGreer, Meisner, Metcalfe, Miller, Miquel, Moustakas, Myers, Naik, Newman, Nichol, Nicola, Nicolati~da Costa, Nie, Niz, Norberg, Nord, Norman, Nugent, O'Brien, Oh, Olsen, Padilla, Padmanabhan, Padmanabhan, Palanque-Delabrouille, Palmese, Pappalardo, P{\^{a}}ris, Park, Patej, Peacock, Peiris, Peng, Percival, Perruchot, Pieri, Pogge, Pollack, Poppett, Prada, Prakash, Probst, Rabinowitz, Raichoor, Ree, Refregier, Regal, Reid, Reil, Rezaie, Rockosi, Roe, Ronayette, Roodman, Ross, Ross, Rossi, Rozo, Ruhlmann-Kleider, Rykoff, Sabiu, Samushia, Sanchez, Sanchez, Schlegel, Schneider, Schubnell, Secroun, Seljak, Seo, Serrano, Shafieloo, Shan, Sharples, Sholl, Shourt, Silber, Silva, Sirk, Slosar, Smith, Smoot, Som, Song, Sprayberry, Staten, Stefanik, Tarle, Sien~Tie, Tinker, Tojeiro, Valdes, Valenzuela, Valluri, Vargas-Magana,
  Verde, Walker, Wang, Wang, Weaver, Weaverdyck, Wechsler, Weinberg, White, Yang, Yeche, Zhang, Zhao, Zheng, Zhou, Zhou, Zhu, Zou, \& Zu}]{DESI2016}
{DESI Collaboration}, Aghamousa, A., Aguilar, J., {et~al.} 2016, arXiv:1611.00036

\bibitem[{Di~Valentino {et~al.}(2021)Di~Valentino, Gariazzo, \& Mena}]{Valentino2021}
Di~Valentino, E., Gariazzo, S., \& Mena, O. 2021, Physical Review D, 104, 083504

\bibitem[{Dodelson \& Schmidt(2020)}]{Dodelson2020}
Dodelson, S. \& Schmidt, F. 2020, Academic Press, Elsevier, 1

\bibitem[{Dvali {et~al.}(2000)Dvali, Gabadadze, \& Porrati}]{Dvali2000}
Dvali, G., Gabadadze, G., \& Porrati, M. 2000, Physics Letters B, 485, 208

\bibitem[{{Euclid Collaboration} {et~al.}(2021){Euclid Collaboration}, Knabenhans, Stadel, Potter, Dakin, Hannestad, Tram, Marelli, Schneider, Teyssier, Fosalba, Andreon, Auricchio, Baccigalupi, Balaguera-Antol{\'{i}}nez, Baldi, Bardelli, Battaglia, Bender, Biviano, Bodendorf, Bozzo, Branchini, Brescia, Burigana, Cabanac, Camera, Capobianco, Cappi, Carbone, Carretero, Carvalho, Casas, Casas, Castellano, Castignani, Cavuoti, Cledassou, Colodro-Conde, Congedo, Conselice, Conversi, Copin, Corcione, Coupon, Courtois, Da~Silva, De~La~Torre, Ferdinando, Duncan, Dupac, Fabbian, Farrens, Ferreira, Finelli, Frailis, Franceschi, Galeotta, Garilli, Giocoli, Gozaliasl, Graci{\'{a}}-Carpio, Grupp, Guzzo, Holmes, Hormuth, Israel, Jahnke, Keihanen, Kermiche, Kirkpatrick, Kubik, Kunz, Kurki-Suonio, Ligori, Lilje, Lloro, Maino, Marggraf, Markovic, Martinet, Marulli, Massey, Mauri, Maurogordato, Medinaceli, Meneghetti, Metcalf, Meylan, Moresco, Morin, Moscardini, Munari, Neissner, Niemi, Padilla, Paltani, Pasian, Patrizii,
  Pettorino, Pires, Polenta, Poncet, Raison, Renzi, Rhodes, Riccio, Romelli, Roncarelli, Saglia, S{\'{a}}nchez, Sapone, Schneider, Scottez, Secroun, Serrano, Sirignano, Sirri, Stanco, Sureau, Cresp{\'{i}}, Taylor, Tenti, Tereno, Toledo-Moreo, Torradeflot, Valenziano, Valiviita, Vassallo, Viel, Wang, Welikala, Whittaker, Zacchei, \& Zucca}]{knabenhans2021}
{Euclid Collaboration}, Knabenhans, M., Stadel, J., {et~al.} 2021, Monthly Notices of the Royal Astronomical Society, 505, 2840

\bibitem[{{Euclid Collaboration} {et~al.}(2022){Euclid Collaboration}, Scaramella, Amiaux, Mellier, Burigana, Carvalho, Cuillandre, Da~Silva, Derosa, Dinis, Maiorano, Maris, Tereno, Laureijs, Boenke, Buenadicha, Dupac, Gaspar~Venancio, G{\'{o}}mez-{\'{A}}lvarez, Hoar, Lorenzo~Alvarez, Racca, Saavedra-Criado, Schwartz, Vavrek, Schirmer, Aussel, Azzollini, Cardone, Cropper, Ealet, Garilli, Gillard, Granett, Guzzo, Hoekstra, Jahnke, Kitching, Maciaszek, Meneghetti, Miller, Nakajima, Niemi, Pasian, Percival, Pottinger, Sauvage, Scodeggio, Wachter, Zacchei, Aghanim, Amara, Auphan, Auricchio, Awan, Balestra, Bender, Bodendorf, Bonino, Branchini, Brau-Nogue, Brescia, Candini, Capobianco, Carbone, Carlberg, Carretero, Casas, Castander, Castellano, Cavuoti, Cimatti, Cledassou, Congedo, Conselice, Conversi, Copin, Corcione, Costille, Courbin, Degaudenzi, Douspis, Dubath, Duncan, Dusini, Farrens, Ferriol, Fosalba, Fourmanoit, Frailis, Franceschi, Franzetti, Fumana, Gillis, Giocoli, Grazian, Grupp, Haugan, Holmes,
  Hormuth, Hudelot, Kermiche, Kiessling, Kilbinger, Kohley, Kubik, K{\"{u}}mmel, Kunz, Kurki-Suonio, Lahav, Ligori, Lilje, Lloro, Mansutti, Marggraf, Markovic, Marulli, Massey, Maurogordato, Melchior, Merlin, Meylan, Mohr, Moresco, Morin, Moscardini, Munari, Nichol, Padilla, Paltani, Peacock, Pedersen, Pettorino, Pires, Poncet, Popa, Pozzetti, Raison, Rebolo, Rhodes, Rix, Roncarelli, Rossetti, Saglia, Schneider, Schrabback, Secroun, Seidel, Serrano, Sirignano, Sirri, Skottfelt, Stanco, Starck, Tallada-Cresp{\'{i}}, Tavagnacco, Taylor, Teplitz, Toledo-Moreo, Torradeflot, Trifoglio, Valentijn, Valenziano, Verdoes~Kleijn, Wang, Welikala, Weller, Wetzstein, Zamorani, Zoubian, Andreon, Baldi, Bardelli, Boucaud, Camera, Di~Ferdinando, Fabbian, Farinelli, Galeotta, Graci{\'{a}}-Carpio, Maino, Medinaceli, Mei, Neissner, Polenta, Renzi, Romelli, Rosset, Sureau, Tenti, Vassallo, Zucca, Baccigalupi, Balaguera-Antol{\'{i}}nez, Battaglia, Biviano, Borgani, Bozzo, Cabanac, Cappi, Casas, Castignani, Colodro-Conde, Coupon,
  Courtois, Cuby, De~La~Torre, Desai, Dole, Fabricius, Farina, Ferreira, Finelli, Flose-Reimberg, Fotopoulou, Ganga, Gozaliasl, Hook, Keihanen, Kirkpatrick, Liebing, Lindholm, Mainetti, Martinelli, Martinet, Maturi, McCracken, Metcalf, Morgante, Nightingale, Nucita, Patrizii, Potter, Riccio, S{\'{a}}nchez, Sapone, Schewtschenko, Schultheis, Scottez, Teyssier, Tutusaus, Valiviita, Viel, Vriend, \& Whittaker}]{Scaramella2022}
{Euclid Collaboration}, Scaramella, R., Amiaux, J., {et~al.} 2022, Astronomy {\&} Astrophysics, 662, A112

\bibitem[{{Euclid Consortium} {et~al.}(2023){Euclid Consortium}, Adamek, Angulo, Arnold, Baldi, Biagetti, Bose, Carbone, Castro, Dakin, Dolag, Elbers, Fidler, Giocoli, Hannestad, Hassani, Hern{\'{a}}ndez-Aguayo, Koyama, Li, Mauland, Monaco, Moretti, Mota, Partmann, Parimbelli, Potter, Schneider, Schulz, Smith, Springel, Stadel, Tram, Viel, Villaescusa-Navarro, Winther, Wright, Zennaro, Aghanim, Amendola, Auricchio, Bonino, Branchini, Brescia, Camera, Capobianco, Cardone, Carretero, Castander, Castellano, Cavuoti, Cimatti, Cledassou, Congedo, Conversi, Copin, Da~Silva, Degaudenzi, Douspis, Dubath, Duncan, Dupac, Dusini, Farrens, Ferriol, Fosalba, Frailis, Franceschi, Galeotta, Garilli, Gillard, Gillis, Grazian, Haugan, Holmes, Hornstrup, Jahnke, Kermiche, Kiessling, Kilbinger, Kitching, Kunz, Kurki-Suonio, Lilje, Lloro, Mansutti, Marggraf, Marulli, Massey, Medinaceli, Meneghetti, Meylan, Moresco, Moscardini, Munari, Niemi, Padilla, Paltani, Pasian, Pedersen, Percival, Pettorino, Polenta, Poncet, Popa, Raison,
  Rebolo, Renzi, Rhodes, Riccio, Romelli, Roncarelli, Saglia, Sapone, Sartoris, Schneider, Schrabback, Secroun, Seidel, Sirignano, Sirri, Stanco, Starck, Tallada-Cresp{\'{i}}, Taylor, Tereno, Toledo-Moreo, Torradeflot, Tutusaus, Valenziano, Vassallo, Wang, Weller, Zacchei, Zamorani, Zoubian, Fabbian, \& Scottez}]{Adamek2023}
{Euclid Consortium}, Adamek, J., Angulo, R.~E., {et~al.} 2023, Journal of Cosmology and Astroparticle Physics, 2023, 035

\bibitem[{Feng {et~al.}(2016)Feng, Chu, Seljak, \& McDonald}]{Feng2016}
Feng, Y., Chu, M.~Y., Seljak, U., \& McDonald, P. 2016, Monthly Notices of the Royal Astronomical Society, 463, 2273

\bibitem[{Fiorini {et~al.}(2022)Fiorini, Koyama, \& Izard}]{Fiorini2022}
Fiorini, B., Koyama, K., \& Izard, A. 2022, Journal of Cosmology and Astroparticle Physics, 2022, 028

\bibitem[{Giblin {et~al.}(2019)Giblin, Cataneo, Moews, \& Heymans}]{Giblin2019}
Giblin, B., Cataneo, M., Moews, B., \& Heymans, C. 2019, Monthly Notices of the Royal Astronomical Society, 490, 4826

\bibitem[{Giocoli {et~al.}(2018)Giocoli, Baldi, \& Moscardini}]{Giocoli2018}
Giocoli, C., Baldi, M., \& Moscardini, L. 2018, Monthly Notices of the Royal Astronomical Society, 481, 2813

\bibitem[{Gupta {et~al.}(2023)Gupta, Hellwing, \& Bilicki}]{Gupta2023}
Gupta, S., Hellwing, W.~A., \& Bilicki, M. 2023, Physical Review D, 107, 083525

\bibitem[{Hannestad {et~al.}(2020)Hannestad, Wong, Thibault, Masson, Kobayashi, Sakka, Suzuki, \& Iop}]{HannestadWong2020}
Hannestad, S., Wong, Y. Y.~Y., Thibault, V., {et~al.} 2020, Journal of Cosmology and Astroparticle Physics, 2020, 028

\bibitem[{Hassani \& Lombriser(2020)}]{Hassani2020}
Hassani, F. \& Lombriser, L. 2020, Monthly Notices of the Royal Astronomical Society, 497, 1885

\bibitem[{Heitmann {et~al.}(2006)Heitmann, Higdon, Nakhleh, \& Habib}]{Heitmann2006}
Heitmann, K., Higdon, D., Nakhleh, C., \& Habib, S. 2006, The Astrophysical Journal, 646, L1

\bibitem[{Heitmann {et~al.}(2013)Heitmann, Lawrence, Kwan, Habib, \& Higdon}]{Heitmann2013}
Heitmann, K., Lawrence, E., Kwan, J., Habib, S., \& Higdon, D. 2013, The Astrophysical Journal, 780, 111

\bibitem[{Hinterbichler {et~al.}(2011)Hinterbichler, Khoury, Levy, \& Matas}]{Hinterbichler2011}
Hinterbichler, K., Khoury, J., Levy, A., \& Matas, A. 2011, Physical Review D - Particles, Fields, Gravitation and Cosmology, 84, 103521

\bibitem[{Hu \& Sawicki(2007)}]{Hu2007}
Hu, W. \& Sawicki, I. 2007, Physical Review D - Particles, Fields, Gravitation and Cosmology, 76, 064004

\bibitem[{Izard {et~al.}(2016)Izard, Crocce, \& Fosalba}]{Izard2016}
Izard, A., Crocce, M., \& Fosalba, P. 2016, Monthly Notices of the Royal Astronomical Society, 459, 2327

\bibitem[{{J-PAS Collaboration} {et~al.}(2014){J-PAS Collaboration}, Benitez, Dupke, Moles, Sodre, Cenarro, Marin-Franch, Taylor, Cristobal, Fernandez-Soto, de~Oliveira, Cepa-Nogue, Abramo, Alcaniz, Overzier, Hernandez-Monteagudo, Alfaro, Kanaan, Carvano, Reis, Gonzalez, Ascaso, Ballesteros, Xavier, Varela, Ederoclite, Ramio, Broadhurst, Cypriano, Angulo, Diego, Zandivarez, Diaz, Melchior, Umetsu, Spinelli, Zitrin, Coe, Yepes, Vielva, Sahni, Marcos-Caballero, Kitaura, Maroto, Masip, Tsujikawa, Carneiro, Nuevo, Carvalho, Reboucas, Carvalho, Abdalla, Bernui, Pigozzo, Ferreira, Devi, Bengaly~Jr., Campista, Amorim, Asari, Bongiovanni, Bonoli, Bruzual, Cardiel, Cava, Fernandes, Coelho, Cortesi, Delgado, Garcia, Espinosa, Galliano, Gonzalez-Serrano, Falcon-Barroso, Fritz, Fernandes, Gorgas, Hoyos, Jimenez-Teja, Lopez-Aguerri, Juan, Mateus, Molino, Novais, OMill, Oteo, Perez-Gonzalez, Poggianti, Proctor, Ricciardelli, Sanchez-Blazquez, Storchi-Bergmann, Telles, Schoennell, Trujillo, Vazdekis, Viironen, Daflon,
  Aparicio-Villegas, Rocha, Ribeiro, Borges, Martins, Marcolino, Martinez-Delgado, Perez-Torres, Siffert, Calvao, Sako, Kessler, Alvarez-Candal, De~Pra, Roig, Lazzaro, Gorosabel, de~Oliveira, Lima-Neto, Irwin, Liu, Alvarez, Balmes, Chueca, Costa-Duarte, da~Costa, Dantas, Diaz, Fabregat, Ferrari, Gavela, Gracia, Gruel, Gutierrez, Guzman, Hernandez-Fernandez, Herranz, Hurtado-Gil, Jablonsky, Laporte, Tiran, Licandro, Lima, Martin, Martinez, Montero, Penteado, Pereira, Peris, Quilis, Sanchez-Portal, Soja, Solano, Torra, \& Valdivielso}]{Benitez2014}
{J-PAS Collaboration}, Benitez, N., Dupke, R., {et~al.} 2014, arXiv:1403.5237

\bibitem[{Joudaki {et~al.}(2022)Joudaki, Ferreira, Lima, \& Winther}]{Joudaki2022}
Joudaki, S., Ferreira, P.~G., Lima, N.~A., \& Winther, H.~A. 2022, Physical Review D, 105, 043522

\bibitem[{{KATRIN Collaboration} {et~al.}(2022){KATRIN Collaboration}, Aker, Beglarian, Behrens, Berlev, Besserer, Bieringer, Block, Bobien, B{\"{o}}ttcher, Bornschein, Bornschein, Brunst, Caldwell, Carney, La~Cascio, Chilingaryan, Choi, Debowski, Deffert, Descher, D{\'{i}}az~Barrero, Doe, Dragoun, Drexlin, Eitel, Ellinger, Engel, Enomoto, Felden, Formaggio, Fr{\"{a}}nkle, Franklin, Friedel, Fulst, Gauda, Gil, Gl{\"{u}}ck, Gr{\"{o}}ssle, Gumbsheimer, Gupta, H{\"{o}}hn, Hannen, Hau{\ss}mann, Helbing, Hickford, Hiller, Hillesheimer, Hinz, Houdy, Huber, Jansen, Karl, Kellerer, Kellerer, Kleifges, Klein, K{\"{o}}hler, K{\"{o}}llenberger, Kopmann, Korzeczek, Koval{\'{i}}k, Krasch, Krause, Kunka, Lasserre, Le, Lebeda, Lehnert, Lokhov, Machatschek, Malcherek, Mark, Marsteller, Martin, Melzer, Menshikov, Mertens, Mostafa, M{\"{u}}ller, Neumann, Niemes, Oelpmann, Parno, Poon, Poyato, Priester, Ramachandran, Robertson, Rodejohann, R{\"{o}}llig, R{\"{o}}ttele, Rodenbeck, Ry{\v{s}}av{\'{y}}, Sack, Saenz, Sch{\"{a}}fer,
  Schaller~n{\'{e}}e Pollithy, Schimpf, Schl{\"{o}}sser, Schl{\"{o}}sser, Schl{\"{u}}ter, Schneidewind, Schrank, Schulz, Schwemmer, {\v{S}}ef{\v{c}}{\'{i}}k, Sibille, Siegmann, Slez{\'{a}}k, Spanier, Steidl, Sturm, Sun, Tcherniakhovski, Telle, Thorne, Th{\"{u}}mmler, Titov, Tkachev, Urban, Valerius, V{\'{e}}nos, Vizcaya~Hern{\'{a}}ndez, Weinheimer, Welte, Wendel, Wilkerson, Wolf, W{\"{u}}stling, Wydra, Xu, Yen, Zadoroghny, \& Zeller}]{Aker2022}
{KATRIN Collaboration}, Aker, M., Beglarian, A., {et~al.} 2022, Nature Physics 2022 18:2, 18, 160

\bibitem[{Khoury \& Weltman(2004{\natexlab{a}})}]{Khoury2004}
Khoury, J. \& Weltman, A. 2004{\natexlab{a}}, Physical Review D, 69, 044026

\bibitem[{Khoury \& Weltman(2004{\natexlab{b}})}]{KhouryWeltman2004}
Khoury, J. \& Weltman, A. 2004{\natexlab{b}}, Physical Review Letters, 93, 171104

\bibitem[{Klypin {et~al.}(2020)Klypin, Prada, \& Byun}]{Klypin2020}
Klypin, A., Prada, F., \& Byun, J. 2020, Monthly Notices of the Royal Astronomical Society, 496, 3862

\bibitem[{Koda {et~al.}(2016)Koda, Blake, Beutler, Kazin, \& Marin}]{Koda2016}
Koda, J., Blake, C., Beutler, F., Kazin, E., \& Marin, F. 2016, Monthly Notices of the Royal Astronomical Society, 459, 2118

\bibitem[{Koyama(2016)}]{Koyama2016}
Koyama, K. 2016, Reports on Progress in Physics, 79, 046902

\bibitem[{Koyama {et~al.}(2009)Koyama, Taruya, \& Hiramatsu}]{Koyama2009}
Koyama, K., Taruya, A., \& Hiramatsu, T. 2009, Physical Review D - Particles, Fields, Gravitation and Cosmology, 79, 123512

\bibitem[{Kwan {et~al.}(2015)Kwan, Heitmann, Habib, Padmanabhan, Lawrence, Finkel, Frontiere, \& Pope}]{Kwan2015}
Kwan, J., Heitmann, K., Habib, S., {et~al.} 2015, The Astrophysical Journal, 810, 35

\bibitem[{Laureijs {et~al.}(2011)Laureijs, Amiaux, Arduini, Augu{\`{e}}res, Brinchmann, Cole, Cropper, Dabin, Duvet, Ealet, Garilli, Gondoin, Guzzo, Hoar, Hoekstra, Holmes, Kitching, Maciaszek, Mellier, Pasian, Percival, Rhodes, Criado, Sauvage, Scaramella, Valenziano, Warren, Bender, Castander, Cimatti, F{\`{e}}vre, Kurki-Suonio, Levi, Lilje, Meylan, Nichol, Pedersen, Popa, Lopez, Rix, Rottgering, Zeilinger, Grupp, Hudelot, Massey, Meneghetti, Miller, Paltani, Paulin-Henriksson, Pires, Saxton, Schrabback, Seidel, Walsh, Aghanim, Amendola, Bartlett, Baccigalupi, Beaulieu, Benabed, Cuby, Elbaz, Fosalba, Gavazzi, Helmi, Hook, Irwin, Kneib, Kunz, Mannucci, Moscardini, Tao, Teyssier, Weller, Zamorani, Osorio, Boulade, Foumond, Di~Giorgio, Guttridge, James, Kemp, Martignac, Spencer, Walton, Bl{\"{u}}mchen, Bonoli, Bortoletto, Cerna, Corcione, Fabron, Jahnke, Ligori, Madrid, Martin, Morgante, Pamplona, Prieto, Riva, Toledo, Trifoglio, Zerbi, Abdalla, Douspis, Grenet, Borgani, Bouwens, Courbin, Delouis, Dubath,
  Fontana, Frailis, Grazian, Koppenh{\"{o}}fer, Mansutti, Melchior, Mignoli, Mohr, Neissner, Noddle, Poncet, Scodeggio, Serrano, Shane, Starck, Surace, Taylor, Verdoes-Kleijn, Vuerli, Williams, Zacchei, Altieri, Sanz, Kohley, Oosterbroek, Astier, Bacon, Bardelli, Baugh, Bellagamba, Benoist, Bianchi, Biviano, Branchini, Carbone, Cardone, Clements, Colombi, Conselice, Cresci, Deacon, Dunlop, Fedeli, Fontanot, Franzetti, Giocoli, Garcia-Bellido, Gow, Heavens, Hewett, Heymans, Holland, Huang, Ilbert, Joachimi, Jennins, Kerins, Kiessling, Kirk, Kotak, Krause, Lahav, van Leeuwen, Lesgourgues, Lombardi, Magliocchetti, Maguire, Majerotto, Maoli, Marulli, Maurogordato, McCracken, McLure, Melchiorri, Merson, Moresco, Nonino, Norberg, Peacock, Pello, Penny, Pettorino, Di~Porto, Pozzetti, Quercellini, Radovich, Rassat, Roche, Ronayette, Rossetti, Sartoris, Schneider, Semboloni, Serjeant, Simpson, Skordis, Smadja, Smartt, Spano, Spiro, Sullivan, Tilquin, Trotta, Verde, Wang, Williger, Zhao, Zoubian, \&
  Zucca}]{Laureijs2011}
Laureijs, R., Amiaux, J., Arduini, S., {et~al.} 2011, arXiv:1110.3193

\bibitem[{Leclercq {et~al.}(2020)Leclercq, Faure, Lavaux, Wandelt, Jaffe, Heavens, \& Percival}]{Leclercq2020}
Leclercq, F., Faure, B., Lavaux, G., {et~al.} 2020, Astronomy {\&} Astrophysics, 639, A91

\bibitem[{Lesgourgues(2011)}]{Lesgourgues2011}
Lesgourgues, J. 2011, arXiv:1104.2932

\bibitem[{Lesgourgues \& Pastor(2006)}]{Lesgourgues2006}
Lesgourgues, J. \& Pastor, S. 2006, Physics Reports, 429, 307

\bibitem[{Li {et~al.}(2012)Li, Zhao, Teyssier, \& Koyama}]{Liecosmog2012}
Li, B., Zhao, G.~B., Teyssier, R., \& Koyama, K. 2012, Journal of Cosmology and Astroparticle Physics, 2012, 051

\bibitem[{Linder(2003)}]{Linder2003}
Linder, E.~V. 2003, Physical Review Letters, 90, 4

\bibitem[{Liu {et~al.}(2018)Liu, Bird, Matilla, Hill, Haiman, Madhavacheril, Spergel, \& Petri}]{Liu2018}
Liu, J., Bird, S., Matilla, J. M.~Z., {et~al.} 2018, Journal of Cosmology and Astroparticle Physics, 2018, 049

\bibitem[{Llinares {et~al.}(2014)Llinares, Mota, \& Winther}]{Llinares2014}
Llinares, C., Mota, D.~F., \& Winther, H.~A. 2014, Astronomy {\&} Astrophysics, 562, A78

\bibitem[{{LSST Collaboration} {et~al.}(2019){LSST Collaboration}, Ivezic, Kahn, Tyson, Abel, Acosta, Allsman, Alonso, AlSayyad, Anderson, Andrew, Angel, Angeli, Ansari, Antilogus, Araujo, Armstrong, Arndt, Astier, Aubourg, Auza, Axelrod, Bard, Barr, Barrau, Bartlett, Bauer, Bauman, Baumont, Bechtol, Bechtol, Becker, Becla, Beldica, Bellavia, Bianco, Biswas, Blanc, Blazek, Blandford, Bloom, Bogart, Bond, Booth, Borgland, Borne, Bosch, Boutigny, Brackett, Bradshaw, Brandt, Brown, Bullock, Burchat, Burke, Cagnoli, Calabrese, Callahan, Callen, Carlin, Carlson, Chandrasekharan, Charles-Emerson, Chesley, Cheu, Chiang, Chiang, Chirino, Chow, Ciardi, Claver, Cohen-Tanugi, Cockrum, Coles, Connolly, Cook, Cooray, Covey, Cribbs, Cui, Cutri, Daly, Daniel, Daruich, Daubard, Daues, Dawson, Delgado, Dellapenna, Peyster, Val-Borro, Digel, Doherty, Dubois, Dubois-Felsmann, Durech, Economou, Eifler, Eracleous, Emmons, Neto, Ferguson, Figueroa, Fisher-Levine, Focke, Foss, Frank, Freemon, Gangler, Gawiser, Geary, Gee, Geha,
  Gessner, Gibson, Gilmore, Glanzman, Glick, Goldina, Goldstein, Goodenow, Graham, Gressler, Gris, Guy, Guyonnet, Haller, Harris, Hascall, Haupt, Hernandez, Herrmann, Hileman, Hoblitt, Hodgson, Hogan, Howard, Huang, Huffer, Ingraham, Innes, Jacoby, Jain, Jammes, Jee, Jenness, Jernigan, Jevremovi{\'{c}}, Johns, Johnson, Johnson, Jones, Juramy-Gilles, Juri{\'{c}}, Kalirai, Kallivayalil, Kalmbach, Kantor, Karst, Kasliwal, Kelly, Kessler, Kinnison, Kirkby, Knox, Kotov, Krabbendam, Krughoff, Kub{\'{a}}nek, Kuczewski, Kulkarni, Ku, Kurita, Lage, Lambert, Lange, Langton, Guillou, Levine, Liang, Lim, Lintott, Long, Lopez, Lotz, Lupton, Lust, MacArthur, Mahabal, Mandelbaum, Markiewicz, Marsh, Marshall, Marshall, May, McKercher, McQueen, Meyers, Migliore, Miller, Mills, Miraval, Moeyens, Moolekamp, Monet, Moniez, Monkewitz, Montgomery, Morrison, Mueller, Muller, Arancibia, Neill, Newbry, Nief, Nomerotski, Nordby, O’Connor, Oliver, Olivier, Olsen, O’Mullane, Ortiz, Osier, Owen, Pain, Palecek, Parejko, Parsons,
  Pease, Peterson, Peterson, Petravick, Petrick, Petry, Pierfederici, Pietrowicz, Pike, Pinto, Plante, Plate, Plutchak, Price, Prouza, Radeka, Rajagopal, Rasmussen, Regnault, Reil, Reiss, Reuter, Ridgway, Riot, Ritz, Robinson, Roby, Roodman, Rosing, Roucelle, Rumore, Russo, Saha, Sassolas, Schalk, Schellart, Schindler, Schmidt, Schneider, Schneider, Schoening, Schumacher, Schwamb, Sebag, Selvy, Sembroski, Seppala, Serio, Serrano, Shaw, Shipsey, Sick, Silvestri, Slater, Smith, Smith, Sobhani, Soldahl, Storrie-Lombardi, Stover, Strauss, Street, Stubbs, Sullivan, Sweeney, Swinbank, Szalay, Takacs, Tether, Thaler, Thayer, Thomas, Thornton, Thukral, Tice, Trilling, Turri, Berg, Berk, Vetter, Virieux, Vucina, Wahl, Walkowicz, Walsh, Walter, Wang, Wang, Warner, Wiecha, Willman, Winters, Wittman, Wolff, Wood-Vasey, Wu, Xin, Yoachim, \& Zhan}]{Ivezic2019}
{LSST Collaboration}, Ivezic, Z., Kahn, S.~M., {et~al.} 2019, The Astrophysical Journal, 873, 111

\bibitem[{Marsh(2016)}]{Marsh2016}
Marsh, D.~J. 2016, Physics Reports, 643, 1

\bibitem[{Martin {et~al.}(2014)Martin, Ringeval, \& Vennin}]{Martin2014}
Martin, J., Ringeval, C., \& Vennin, V. 2014, Physics of the Dark Universe, 5, 75

\bibitem[{Mauland {et~al.}(2023)Mauland, Elgar{\o}y, Mota, \& Winther}]{Mauland2023}
Mauland, R., Elgar{\o}y, {\O}., Mota, D.~F., \& Winther, H.~A. 2023, Astronomy {\&} Astrophysics, 674, A185

\bibitem[{McKay {et~al.}(1979)McKay, Beckman, \& Conover}]{McKay1979}
McKay, M.~D., Beckman, R.~J., \& Conover, W.~J. 1979, Technometrics, 21, 239

\bibitem[{Moran {et~al.}(2023)Moran, Heitmann, Lawrence, Habib, Bingham, Upadhye, Kwan, Higdon, \& Payne}]{Moran2023}
Moran, K.~R., Heitmann, K., Lawrence, E., {et~al.} 2023, Monthly Notices of the Royal Astronomical Society, 520, 3443

\bibitem[{Nishimichi {et~al.}(2019)Nishimichi, Takada, Takahashi, Osato, Shirasaki, Oogi, Miyatake, Oguri, Murata, Kobayashi, \& Yoshida}]{Nishimichi2019}
Nishimichi, T., Takada, M., Takahashi, R., {et~al.} 2019, The Astrophysical Journal, 884, 29

\bibitem[{{Particle Data Group} {et~al.}(2022){Particle Data Group}, Workman, \& {Others}}]{PDG2022}
{Particle Data Group}, Workman, R.~L., \& {Others}. 2022, PTEP, 2022, 083C01

\bibitem[{Partmann {et~al.}(2020)Partmann, Fidler, Rampf, \& Hahn}]{Partmann2020}
Partmann, C., Fidler, C., Rampf, C., \& Hahn, O. 2020, Journal of Cosmology and Astroparticle Physics, 2020, 018

\bibitem[{Peebles(1980)}]{Peebles1980}
Peebles, E. P.~J. 1980, Princeton University Press

\bibitem[{{Planck Collaboration} {et~al.}(2020){Planck Collaboration}, Aghanim, Akrami, Ashdown, Aumont, Baccigalupi, Ballardini, Banday, Barreiro, Bartolo, Basak, Battye, Benabed, Bernard, Bersanelli, Bielewicz, Bock, Bond, Borrill, Bouchet, Boulanger, Bucher, Burigana, Butler, Calabrese, Cardoso, Carron, Challinor, Chiang, Chluba, Colombo, Combet, Contreras, Crill, Cuttaia, De~Bernardis, De~Zotti, Delabrouille, Delouis, Di~Valentino, Diego, Dor{\'{e}}, Douspis, Ducout, Dupac, Dusini, Efstathiou, Elsner, En{\ss}lin, Eriksen, Fantaye, Farhang, Fergusson, Fernandez-Cobos, Finelli, Forastieri, Frailis, Fraisse, Franceschi, Frolov, Galeotta, Galli, Ganga, G{\'{e}}nova-Santos, Gerbino, Ghosh, Gonz{\'{a}}lez-Nuevo, G{\'{o}}rski, Gratton, Gruppuso, Gudmundsson, Hamann, Handley, Hansen, Herranz, Hildebrandt, Hivon, Huang, Jaffe, Jones, Karakci, Keih{\"{a}}nen, Keskitalo, Kiiveri, Kim, Kisner, Knox, Krachmalnicoff, Kunz, Kurki-Suonio, Lagache, Lamarre, Lasenby, Lattanzi, Lawrence, Le~Jeune, Lemos, Lesgourgues,
  Levrier, Lewis, Liguori, Lilje, Lilley, Lindholm, L{\'{o}}pez-Caniego, Lubin, Ma, Maci{\'{a}}s-P{\'{e}}rez, Maggio, Maino, Mandolesi, Mangilli, Marcos-Caballero, Maris, Martin, Martinelli, Mart{\'{i}}nez-Gonz{\'{a}}lez, Matarrese, Mauri, McEwen, Meinhold, Melchiorri, Mennella, Migliaccio, Millea, Mitra, Miville-Desch{\^{e}}nes, Molinari, Montier, Morgante, Moss, Natoli, N{\o}rgaard-Nielsen, Pagano, Paoletti, Partridge, Patanchon, Peiris, Perrotta, Pettorino, Piacentini, Polastri, Polenta, Puget, Rachen, Reinecke, Remazeilles, Renzi, Rocha, Rosset, Roudier, Rubi{\~{n}}o-Mart{\'{i}}n, Ruiz-Granados, Salvati, Sandri, Savelainen, Scott, Shellard, Sirignano, Sirri, Spencer, Sunyaev, Suur-Uski, Tauber, Tavagnacco, Tenti, Toffolatti, Tomasi, Trombetti, Valenziano, Valiviita, Van~Tent, Vibert, Vielva, Villa, Vittorio, Wandelt, Wehus, White, White, Zacchei, \& Zonca}]{Planck2018}
{Planck Collaboration}, Aghanim, N., Akrami, Y., {et~al.} 2020, Astronomy {\&} Astrophysics, 641, A6

\bibitem[{Pogosian \& Silvestri(2008)}]{Pogosian2008}
Pogosian, L. \& Silvestri, A. 2008, Physical Review D - Particles, Fields, Gravitation and Cosmology, 77, 023503

\bibitem[{Potter {et~al.}(2017)Potter, Stadel, \& Teyssier}]{Potter2016}
Potter, D., Stadel, J., \& Teyssier, R. 2017, Computational Astrophysics and Cosmology, 4

\bibitem[{Puchwein {et~al.}(2013)Puchwein, Baldi, \& Springel}]{Puchwein2013}
Puchwein, E., Baldi, M., \& Springel, V. 2013, Monthly Notices of the Royal Astronomical Society, 436, 348

\bibitem[{Ramachandra {et~al.}(2021)Ramachandra, Valogiannis, Ishak, \& Heitmann}]{Ramachandra2021}
Ramachandra, N., Valogiannis, G., Ishak, M., \& Heitmann, K. 2021, Physical Review D, 103, 123525

\bibitem[{Ruan {et~al.}(2022)Ruan, Hern{\'{a}}ndez-Aguayo, Li, Arnold, Baugh, Klypin, \& Prada}]{Ruan2022}
Ruan, C.~Z., Hern{\'{a}}ndez-Aguayo, C., Li, B., {et~al.} 2022, Journal of Cosmology and Astroparticle Physics, 2022, 018

\bibitem[{S{\'{a}}ez-Casares {et~al.}(2024)S{\'{a}}ez-Casares, Rasera, \& Li}]{saez-casares2023}
S{\'{a}}ez-Casares, I., Rasera, Y., \& Li, B. 2024, Monthly Notices of the Royal Astronomical Society, 527, 7242

\bibitem[{Savitzky \& Golay(1964)}]{Savitzky1964}
Savitzky, A. \& Golay, M.~J. 1964, Analytical Chemistry, 36, 1627

\bibitem[{Song {et~al.}(2007)Song, Hu, \& Sawicki}]{Song2007}
Song, Y.~S., Hu, W., \& Sawicki, I. 2007, Physical Review D - Particles, Fields, Gravitation and Cosmology, 75, 044004

\bibitem[{Sotiriou \& Faraoni(2010)}]{Sotiriou2010}
Sotiriou, T.~P. \& Faraoni, V. 2010, Reviews of Modern Physics, 82, 451

\bibitem[{Springel {et~al.}(2021)Springel, Pakmor, Zier, \& Reinecke}]{Springel2021}
Springel, V., Pakmor, R., Zier, O., \& Reinecke, M. 2021, Monthly Notices of the Royal Astronomical Society, 506, 2871

\bibitem[{Tassev {et~al.}(2015)Tassev, Eisenstein, Wandelt, \& Zaldarriaga}]{Tassev2015}
Tassev, S., Eisenstein, D.~J., Wandelt, B.~D., \& Zaldarriaga, M. 2015, arXiv:1502.07751

\bibitem[{Tassev {et~al.}(2013)Tassev, Zaldarriaga, \& Eisenstein}]{Tassev2013}
Tassev, S., Zaldarriaga, M., \& Eisenstein, D.~J. 2013, Journal of Cosmology and Astroparticle Physics, 2013, 036

\bibitem[{Thomson(2013)}]{Thomson2013}
Thomson, M. 2013, Cambridge University press

\bibitem[{Valogiannis \& Bean(2017)}]{Valogiannis2017}
Valogiannis, G. \& Bean, R. 2017, Physical Review D, 95, 103515

\bibitem[{Villaescusa-Navarro {et~al.}(2018)Villaescusa-Navarro, Naess, Genel, Pontzen, Wandelt, Anderson, Font-Ribera, Battaglia, \& Spergel}]{Villaescusa2018}
Villaescusa-Navarro, F., Naess, S., Genel, S., {et~al.} 2018, The Astrophysical Journal, 867, 137

\bibitem[{Weinberger {et~al.}(2020)Weinberger, Springel, \& Pakmor}]{Weinberger2020}
Weinberger, R., Springel, V., \& Pakmor, R. 2020, The Astrophysical Journal Supplement Series, 248, 32

\bibitem[{Wetterich(1988)}]{Wetterich1988}
Wetterich, C. 1988, Nuclear Physics B, 302, 668

\bibitem[{Will(2014)}]{Will2014}
Will, C.~M. 2014, Living Reviews in Relativity 2014 17:1, 17, 1

\bibitem[{Winther {et~al.}(2019)Winther, Casas, Baldi, Koyama, Li, Lombriser, \& Zhao}]{Winther2019}
Winther, H.~A., Casas, S., Baldi, M., {et~al.} 2019, Physical Review D, 100, 123540

\bibitem[{Winther \& Ferreira(2015)}]{WintherFerreira2015}
Winther, H.~A. \& Ferreira, P.~G. 2015, Physical Review D - Particles, Fields, Gravitation and Cosmology, 91, 123507

\bibitem[{Winther {et~al.}(2017)Winther, Koyama, Manera, Wright, \& Zhao}]{Winther2017}
Winther, H.~A., Koyama, K., Manera, M., Wright, B.~S., \& Zhao, G.~B. 2017, Journal of Cosmology and Astroparticle Physics, 2017, 006

\bibitem[{Winther {et~al.}(2015)Winther, Schmidt, Barreira, Arnold, Bose, Llinares, Baldi, Falck, Hellwing, Koyama, Li, Mota, Puchwein, Smith, \& Zhao}]{Winther2015}
Winther, H.~A., Schmidt, F., Barreira, A., {et~al.} 2015, Monthly Notices of the Royal Astronomical Society, 454, 4208

\bibitem[{Wright {et~al.}(2023)Wright, Sen~Gupta, Baker, Valogiannis, \& Fiorini}]{Wright2023}
Wright, B.~S., Sen~Gupta, A., Baker, T., Valogiannis, G., \& Fiorini, B. 2023, Journal of Cosmology and Astroparticle Physics, 2023, 040

\bibitem[{Wright {et~al.}(2017)Wright, Winther, \& Koyama}]{Wright2017}
Wright, B.~S., Winther, H.~A., \& Koyama, K. 2017, Journal of Cosmology and Astroparticle Physics, 2017, 054

\bibitem[{Zhao(2014)}]{Zhao2014}
Zhao, G.~B. 2014, The Astrophysical Journal Supplement Series, 211, 23

\end{thebibliography}

\end{document}